\documentstyle[12 pt,epsfig]{article}
\begin{document}
\def\emg{electromagnetic}

\def\in{interaction}

\def\ins{interactions}
\begin{center}
{\Large\bf Unitary and Analytic Models in Particle Physics
Phenomenology\footnote{\it Contribution presented at the
International Hadron Structure 2000 Conference held on October 2-7,
2000 in Star\'a Lesn\'a, High Tatra Mountains, Slovak Republic}\\}
\end{center}

\begin{center}
S.Dubni\v cka\\
 Institute of Physics, Slovak Academy of Sciences,
D\'ubravsk\'a cesta 9,
842 28 Bratislava, Slovak Republic\\
\end{center}

\begin{center}
A.Z. Dubni\v ckov\'a\\
 Dept. of Theor. Physics, Comenius University, Mlynsk\'a dolina,
842 48 Bratislava, Slovak Republic\\
\end{center}

\begin{center}
Pavol Str\'{\i}\v zenec\\
Institute of Exp. Physics, Slovak Academy of Sciences, Ko\v sice,
Slovak Republic\\
\end{center}


\abstract{\rm

After two decades of a development of the unitary and analytic
models of the \emg~structure of hadrons and nuclei their main
principles are briefly formulated, then a general scheme of their
applications to the \emg, weak and strong \in~processes are traced
out and finally, some results of their successful applications are
reviewed.}

\section{Introduction}

Hadrons are compound of quarks and as a result, they manifest a
structure in \emg,
 weak and strong \ins~which makes complications at a description of various
processes with hadrons by standard quantum-field theoretical
approaches.

The latter problem in \emg~and weak \ins~of hadrons was
solved by an introduction of functions of one variable, so-called
form factors (FF's) of hadrons, being coefficients in a
decomposition of the unknown hadronic matrix element of the
corresponding current into a maximum number of linearly independent
covariants constructed of four-momenta and spin parameters of the
hadron under consideration. Their behaviour is
 expected to be predicted by the fundamental
theory of quark-gluon \ins, QCD. However, we know, that merely at
sufficiently small distances (thanks to its asymptotic freedom) QCD
becomes a weakly coupled quark-gluon theory to be amenable to a
perturbative expansion in the running coupling constant $\alpha_s$
and predicts just the asymptotic behaviour of FF's. In low momentum
transfer region, where $\alpha_s$ becomes large, the quark-gluon
perturbation theory breaks down and non-perturbative methods in QCD
are not well worked out to give interesting results on the FF's of
hadrons. The same is valid also for the low energy time-like region
where FF's acquire the most complicated, resonant, behaviour. In
this contribution we would like to present the phenomenological
unitary and analytic approach which compensates abovementioned
problems to some extent and renders possible to achieve a line of
interesting results.

\section{Subject of our interest}

The effectuality of our phenomenological approach will be
demonstrated mainly on binary processes caused by \emg, weak and
strong \ins, though also weak decays of $\tau$-lepton into two
pseudoscalar mesons are useful to be considered.
 The worked out unitary and analytic models depend on some number of unknown parameters,
which can be evaluated practically in a comparison of the model under
consideration with a corresponding experimental information. Therefore first,
we concentrate our attention to the {\it\emg} processes of the
type $e^+e^-\to h\bar{h}$ and $e^-h\to e^-h$, $h=P(\pi,K),
B(N,\Lambda,\Sigma, \Xi)$, where on $\pi,K$ and $N$ the abundant experimental
information exists. These processes are described by the cross-sections
\begin{equation}
\sigma^{\rm c.m.}(e^+e^-\to P\bar{P}) = \frac{\pi\alpha^2}{3t}\beta^3_P(t)|F_p(t)|^2
\end{equation}
\begin{equation}
\sigma^{\rm c.m.}(e^+e^-\to B\bar{B}) = \frac{4\pi\alpha^2}{3t}\beta_B(t)
\left[|G^B_M(t)|^2 + \frac{2m_B^2}{t}|G^B_E(t)|^2\right]
\end{equation}
and
\begin{eqnarray}
\frac{{\rm d}\sigma^L(e^-P\to e
^-P)}{{\rm d}\Omega} &=&\frac{\alpha^2}{4E^2}\cdot\frac{{\rm cos}^2\theta/2}{{\rm sin}^4\theta/2}
\cdot\frac{1}{1+\frac{2E}{m_P}{\rm sin}^2\theta/2}\cdot F^2_p(t) \nonumber \\
\frac{{\rm d}\sigma^L(e^-B\to e^-B)}{{\rm d}\Omega} &=&
\frac{\alpha^2}{4E^2}\cdot\frac{{\rm cos}^2\theta/2}{{\rm sin}^4\theta/2}
\cdot\frac{1}{1+\frac{2E}{m_B}{\rm sin}^2\theta/2}\cdot \\
 & &\cdot\left[(F^B_1(t))^2-\frac{t}{4m^2_B}(F^B_2(t))^2-2\frac{t}{4m^2_B}
(F^B_1(t)+F^B_2(t))^2{\rm tg}^2\theta/2\right] \nonumber
\end{eqnarray}
with $\alpha$-the fine structure constant, $\beta_{P,B}(t)=\sqrt{
\frac{t-4m^2_{P,B}}{t}}$; $t=-Q^2$ momentum transfer squared, where
\begin{eqnarray}
G^B_E(t) &=& F^B_1(t)+\frac{t}{4m^2_B}F^B_2(t) \nonumber \\
& & \\
G^B_M(t) &=& F^B_1(t)+F^B_2(t). \nonumber
\end{eqnarray}

All dynamics of the processes $e^+e^-\to h\bar{h}$ and $e^-h\to
e^-h$ is comprised in the functions of one variable $F_P(t)$,
$F_1(t)$, $F_2(t)$, $G^B_E(t)$, $G^B_M(t)$, the so-called \emg~form
factors (FF's) of hadrons.

Next we shall be interested in the {\it weak} processes of the type
$\bar{\nu}_ee^-\to~P^0P^-$   $(P=\pi, K)$ and $\tau^-\to\nu_\tau
P^0P^-$, described by the cross-section
\begin{equation}
\sigma(\bar{\nu}_ee^-\to P^0P^-) = \frac{G^2_F}{48\pi}\;t\;\beta^3_P\;|F^W_P(t)|^2
\label{weakcc}
\end{equation}
and the partial decay width expression
\begin{equation}
\Gamma(\tau^-\to\nu_\tau P^0P^-) =
\frac{G^2_F{\rm cos}^2\Theta_c}{384\pi^3m^3_\tau C^2_p}
\int^{m^2_\tau}_{4m^2_\pi} {\rm d}t(m^2_\tau - t)(m^2_\tau + 2t)
\beta^3_P|F^W_P(t)|^2,
\label{weakpdw}
\end{equation}
respectively, where $G_F$ is the Fermi coupling constant of weak
\ins, $\Theta_c\approx 13^o$ is the Cabibbo angle, $C_\pi=\sqrt{2}$
and $C_K=2$. Again all dynamics of the processes $\bar{\nu_e}e^-\to
P^0P^-$ and $\tau^-\to\nu_\tau P^0P^-$ is comprised in the weak
FF's $F_P^W(t)$.\ Finally, one can consider also {\it strong}
binary processes of the type $a+a\to a+a$; $a+b\to a+b$ and $a+b\to
c+d$ to be described generally by the cross-section
\begin{equation}
\frac{{\rm d}\sigma^{\rm c.m.}(a+b\to c+d)}{{\rm d}\Omega}
 = \frac{\pi^2}{4s}\frac{q}{p}
\frac{1}{(2S_a+1)(2S_b+1)}\sum_{\lambda_a \lambda_b \lambda_c \lambda_d}
|T^{(s)}_{cd;ab}(s,t)|^2,
\label{ccs}
\end{equation}
where $T^{(s)}_{cd;ab}(s,t)$ is the $s$-channel c.m. helicity amplitude,
$p=|\vec{p_a}|=|\vec{p_b}|$; $q=|\vec{p_c}|=|\vec{p_d}|$; $s$-is total
c.m. energy squared, $S_a,S_b,S_c,S_d$ are spins and $\lambda_a$, $\lambda_b$, $\lambda_c$,
 $\lambda_d$ spin projections of $a,b,c,d$ particles, respectively.
If elastic scattering of two spin-zero particles is considered and a
redefinition of the amplitude
\begin{equation}
T(s,t) = -\frac{1}{4\pi^2}M(s,t)
\end{equation}
is taken into account, equation (\ref{ccs}) becomes
\begin{equation}
\frac{{\rm d}\sigma}{{\rm d}\Omega} = \frac{1}{64\pi^2s}|M(s,t)|^2.
\end{equation}
Then one can define the total cross-section by the relation
\begin{equation}
\sigma^{(s)}_{\rm tot}=\int\frac{{\rm d}\sigma}{{\rm d}\Omega} {\rm d}\Omega .
\end{equation}
As a consequence of the conservation of angular momentum, the scattering
amplitude $M(s,t)$ can be expanded in a series of partial wave amplitudes
$f_l(s)$ as functions of only one variable
\begin{equation}
M(s,t)=\sum_{l=0}^\infty (2l+1)f_l(s)P_l(\cos\theta) \label{mexp}
\end{equation}
where $\theta$ is the c.m. scattering angle and $P_l(\cos\theta)$
are the Legendre polynomials. For a general case of particles with
non-zero spin and unequal masses the expansion of helicity
amplitudes in (\ref{ccs}) is exactly analogous to eq.
(\ref{mexp}), except that $l$ is replaced by the total angular
momentum $J$ and the Legendre polynomials $P_l(\cos\theta)$ by the
rotation Wigner's functions $d^J_{\lambda\mu}(\theta)$. Here all
dynamics of the processes $a+a\to a+a$, $a+b\to a+b$ and $a+b\to
c+d$ is comprised in the partial wave amplitudes $f_l(s)$. The eq.
(\ref{mexp}) has enormous practical importance as at low energies
only low partial waves give non negligible contributions so that
the scattering amplitude may be well approximated by a severely
truncated partial wave series.

\section{Common features of considered \emg, weak and strong processes}

One could notice that all considered \emg, weak and strong
\in~processes are completely described provided that the
corresponding functions (\emg~and weak FF's and the partial wave
amplitudes) of one variable are known explicitly. Here we review
their three basic properties, which enable us then to construct for
them realistic phenomenological models.
\begin{itemize}
\item[i\raisebox{-1mm}{)}]
Experimentally measured total cross-sections (in $\tau$-decay it is invariant
mass distribution) as functions of the energy manifest sharp variations
associated with an occurrence of unstable objects, resonances, characterized by
the mass $m_r$ and the width $\Gamma_r$, where the latter is related to the
life-time $\tau_r$ of the resonance by the relation
$$
1/\Gamma_r \approx \tau_r \sim 10^{-23}{\rm s}.
$$
This property is automatically transferred into the absolute values of
corresponding FF's and partial wave amplitudes, and in a construction
of models for the latter functions of one variable this experimental fact
has to be taken into account in the first place.
\item[ii\raisebox{-1mm}{)}]
Electromagnetic and weak FF's, as well as partial wave amplitudes
of strong \in~processes, are analytic functions in the whole
complex plane of their argument, except for isolated singularities.
In the case of \emg~and weak FF's of hadrons they are branch points
on the positive real axis corresponding to {\it normal} (given by
the unitarity condition) and {\it anomalous} (given by allowed
triangle diagrams) thresholds generating many-sheeted Riemann
surface. The first sheet is called physical, all other sheets of
the Riemann surface are unphysical. In the case of partial wave
amplitudes of strong interaction binary processes the situation is
even more complex and depends on the fact, if there is scattering
of particles of
\begin{itemize}
\item[$\bullet$]
equal masses ($a+a\to a+a$):\\
\hspace*{1cm} The
cuts on the positive real axis are similar as for
\emg~and weak FF's. Besides the latter there is also left-hand cut from
$-\infty$ to $s=0$.
\item[$\bullet\bullet$]
unequal masses ($a+b\to a+b$): $m_b > m_a$\\
\hspace*{1cm} On the positive real axis there are unitary cuts similar with
those of \emg~and weak FF's. Besides the latter there is the short cut
between $(m^2_b-m^2_a/m^2_b)$ and $(m^2_b+2m^2_a)$, left-hand cut from
$-\infty$ to $(m_b-m_a)^2$ and circular cut with a center at the origin
and the radius equal to $(m^2_b-m^2_a)$.
\item[$\bullet\bullet\bullet$]
4 different particles ($a+b\to c+d$): \\
\hspace*{1cm} The situation is very complicated and it is out of the scope
of our interest.
\end{itemize}
 Generally, the partial wave amplitude can be written as a quotient of
 two functions
\begin{equation}
f_l(s) = N_l(s)/D_l(s)
\end{equation}
in such a way that $D_l(s)$ contains only the right-hand unitary
cuts and $N_l(s)$ only the left-hand cuts of $f_l(s)$.
 Any $s$-channel resonance in \emg~and weak FF's, as well as in the
 partial wave amplitudes of strong \in~processes, is always associated
 with a pair of complex conjugate poles on unphysical sheets to be generated
 by the branch points on the positive real axis of the $s$-plane.
\item[iii\raisebox{-1mm}{)}]
Asymptotic behaviour of \emg~and weak FF's is related to the number of
constituent quarks $n_q$ of the hadron under consideration by means
of the expression
\begin{equation}
F_h(t)|_{|t|\to\infty}\sim t^{n_q-1}
\end{equation}
i.e.
\begin{equation}
F_P(t)|_{|t|\to\infty}\sim \frac{1}{t}; \;\;\;\;\;\;\;\;\;\;
G^B_{E,M}(t)|_{|t|\to\infty}\sim \frac{1}{t^2}.
\end{equation}
The cross-section of the process $a+b\to a+b$ behaves asymptotically
as follows
\begin{equation}
\frac{{\rm d}\sigma(a+b\to a+b)}{{\rm d}t}|
_{s\to\infty}\sim \frac{1}{s^n}f(t/s)
\end{equation}
where $n=2(n_a+n_b-1)$ and $n_a,n_b$ are numbers of the constituent
quarks of $a,b$ particles.
\end{itemize}

\section{The main idea of construction of unitary and analytic models}
\label{secmainidea}

In the previous section on the base of observation of the experimentally
measured total cross-sections we have been driven to the idea that all
\emg~and weak FF's of hadrons are formed by a contribution of unstable
resonant states, completely described by $m_r$, $\Gamma_r$ and residuum,
and a smooth background generated by unitary cuts on the positive real axis.
In the case of the partial wave amplitudes a smooth background of the latter
type is enlarged by $N_l(s)$ contributions of the left-hand cuts.

Therefore, now, it is straightforward to demonstrate a procedure
how to unify consistently the experimental fact of a creation of
resonant states together with the assumed analytic properties and
the asymptotic behaviour into one analytic function defined on
four-sheeted Riemann surface with complex conjugate pairs of poles,
corresponding to resonances, placed on unphysical sheets. For a
simplicity, first we restrict ourselves to the \emg~FF's of
pseudoscalar mesons $P$. In this case any resonance can be
represented by the diagram (see above)
  \begin{figure}[t]
  \begin{center}
   \epsfig{file=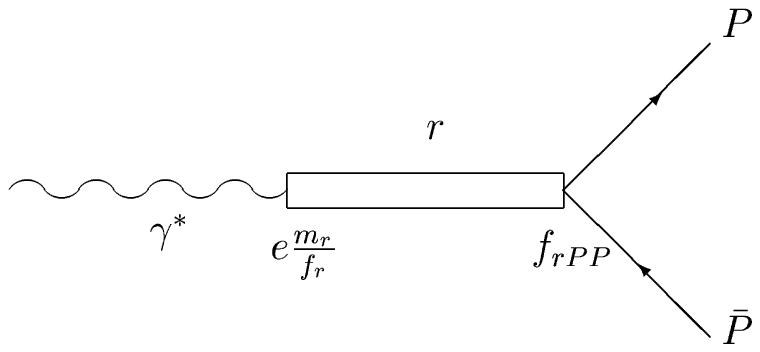,width=0.5\textwidth}
    \end{center}
    \end{figure}
to which the following pole expression
\begin{equation}
\frac{m^2_r}{m^2_r-t}\;(f_{rPP}/f_r)
\label{pole}
\end{equation}
corresponds, where $f_{rPP}$ is the coupling constant of the
resonance to the pseudoscalar meson $P$ and $f_r$ is the  coupling constant
describing the photon-vector-meson transition. The pole expression
(\ref{pole}) has the asymptotics $\sim 1/t$. In
order to transform (\ref{pole}) into one analytic function
\begin{itemize}
\item[i\raisebox{-1mm}{)}]
with two square-root branch points on the positive real axis
\item[ii\raisebox{-1mm}{)}]
with two pairs of complex conjugate poles on unphysical sheets corresponding
to the resonance $r$
\end{itemize}
one proceeds as follows:
\begin{itemize}
\item[{\normalfont\bfseries \textendash}]
first, the nonlinear transformation
\begin{equation}
t = t_0 + \frac{4(t_{in}-t_0)}{[1/W(t)-W(t)]^2}
\label{texp}
\end{equation}
with $t_0$~-~the square-root branch point corresponding to the
lowest threshold and $t_{in}$~-~an effective square-root branch
point simulating contributions of all other relevant thresholds
given by the unitarity condition is applied, which automatically
generates the relations
\begin{equation}
m^2_r = t_0 + \frac{4(t_{in}-t_0)}{[1/W_{r0}-W_{r0}]^2}
\label{m2exp}
\end{equation}
and
\begin{equation}
0 = t_0 + \frac{4(t_{in}-t_0)}{[1/W_N-W_N]^2}
\label{0exp}
\end{equation}
\item[{\normalfont\bfseries \textendash}]
then relations between $W_{r0}$ and $W^*_{r0}$ are utilized
\item[{\normalfont\bfseries \textendash}]
and finally, the instability of the resonance is introduced
by its non-zero width $\Gamma_r \neq 0$.
\end{itemize}

The application of the (\ref{texp})-(\ref{0exp}) to (\ref{pole})
leads to the following factorized form
\begin{eqnarray}
\frac{m^2_r}{m^2_r-t}(f_{rPP}/f_r) &=&\left(\frac{1-W^2}{1-W^2_N}\right)^2\cdot
\label{basefact} \\
& &\cdot\frac{(W_N-W_{r0})(W_N+W_{r0})(W_N-1/W_{r0})(W_N+1/W_{r0})}
{(W-W_{r0})(W+W_{r0})(W-1/W_{r0})(W+1/W_{r0})}\cdot \nonumber \\
& &\cdot(f_{rPP}/f_r) \nonumber
\end{eqnarray}
with asymptotic term (the first term, completely determining the
asymptotic behaviour of (\ref{pole})) and on the so-called
finite-energy term (for $|t|\to\infty$ it turns out to be a real
constant) giving a resonant behaviour around $t=m^2_r$. One can
prove:
\begin{eqnarray}
{\rm a\raisebox{-1mm}{)}~~~if~~} m^2_r-\Gamma^2_r/4 < t_{in} &\Rightarrow& W_{r0} = - W^*_{r0} \nonumber \\
 & & \\
{\rm b\raisebox{-1mm}{)}~~~if~~} m^2_r-\Gamma^2_r/4 > t_{in} &\Rightarrow&  W_{r0} =  1/W^*_{r0} \nonumber
\end{eqnarray}
which lead the eq. (\ref{basefact}) in the case a\raisebox{-1mm}{)}
to the expression
\begin{eqnarray}
\frac{m^2_r}{m^2_r-t}(f_{rPP}/f_r) &=&\left(\frac{1-W^2}{1-W^2_N}\right)^2\cdot
 \label{afact} \\
& &\cdot\frac{(W_N-W_{r0})(W_N-W^*_{r0})(W_N-1/W_{r0})(W_N-1/W^*_{r0})}
{(W-W_{r0})(W-W^*_{r0})(W-1/W_{r0})(W-1/W^*_{r0})}\cdot\nonumber \\
& &\cdot (f_{rPP}/f_r) \nonumber
\end{eqnarray}
and in the case b\raisebox{-1mm}{)} to the following expression
\begin{eqnarray}
\frac{m^2_r}{m^2_r-t}(f_{rPP}/f_r) &=&\left(\frac{1-W^2}{1-W^2_N}\right)^2\cdot
 \label{bfact} \\
& &\cdot\frac{(W_N-W_{r0})(W_N-W^*_{r0})(W_N+W_{r0})(W_N+W^*_{r0})}
{(W-W_{r0})(W-W^*_{r0})(W+W_{r0})(W+W^*_{r0})}\cdot\nonumber\\
& &\cdot (f_{rPP}/f_r). \nonumber
\end{eqnarray}
Lastly, introducing the non-zero width of the resonance by a substitution
\begin{equation}
m^2_r\to (m_r-\Gamma_r/2)^2
\end{equation}
i.e. simply one has to get rid of "0" in sub-indices of
(\ref{afact}) and
(\ref{bfact}), one gets:\\[2mm]
in a\raisebox{-1mm}{)} case\\
\begin{eqnarray}
\frac{m^2_r}{m^2_r-t}(f_{rPP}/f_r) &\to&
\left(\frac{1-W^2}{1-W^2_N}\right)^2\cdot
 \label{aafact} \\
& &\cdot\frac{(W_N-W_r)(W_N-W^*_r)(W_N-1/W_r)(W_N-1/W^*_r)}
{(W-W_r)(W-W^*_r)(W-1/W_r)(W-1/W^*_r)}
\cdot\nonumber\\
& &\cdot
(f_{rPP}/f_r) =  \nonumber\\
& &= \left(\frac{1-W^2}{1-W^2_N}\right)^2 L(W_r)(f_{rPP}/f_r)\nonumber
\end{eqnarray}
and in b\raisebox{-1mm}{)} case\\
\begin{eqnarray}
\frac{m^2_r}{m^2_r-t}(f_{rPP}/f_r) &\to&
\left(\frac{1-W^2}{1-W^2_N}\right)^2\cdot
 \label{bbfact} \\
& &\cdot\frac{(W_N-W_r)(W_N-W^*_r)(W_N+W_r)(W_N+W^*_r)}
{(W-W_r)(W-W^*_r)(W+W_r)(W+W^*_r)}\cdot\nonumber\\
& &\cdot(f_{rPP}/f_r) = \nonumber\\
& &= \left(\frac{1-W^2}{1-W^2_N}\right)^2 H(W_r)(f_{rPP}/f_r)\nonumber
\end{eqnarray}
where no more equality can be used in (\ref{aafact}) and (\ref{bbfact})
between the pole-term and the transformed expressions.

The latter are then analytic functions defined on four-sheeted Riemann
surface, what can be seen explicitly by the inverse transformation
to (\ref{texp})
\begin{equation}
W(t) = i\frac{\sqrt{\left(\frac{t_{in}-t_0}{t_0}\right)^{1/2} +
                         \left(\frac{t-t_0}{t_0}\right)^{1/2}} -
           \sqrt{\left(\frac{t_{in}-t_0}{t_0}\right)^{1/2} -
                         \left(\frac{t-t_0}{t_0}\right)^{1/2}}}
                  {\sqrt{\left(\frac{t_{in}-t_0}{t_0}\right)^{1/2} +
                         \left(\frac{t-t_0}{t_0}\right)^{1/2}} +
           \sqrt{\left(\frac{t_{in}-t_0}{t_0}\right)^{1/2} -
                         \left(\frac{t-t_0}{t_0}\right)^{1/2}}}.
\end{equation}
These expressions have two pairs of complex conjugate poles on:
\begin{itemize}
\item[a\raisebox{-1mm}{)}]
case (\ref{aafact})~-~the second and fourth sheets
\item[b\raisebox{-1mm}{)}]
case (\ref{bbfact})~-~the third and fourth sheets,
\end{itemize}
describing always one resonance under consideration, and the
asymptotic behaviour $\sim 1/t$, to be completely given by the
asymptotic term $\left(\frac{1-W^2}{1-W^2_N}\right)^2$, more
specifically, by its power "2". Here we would like also to note,
that expressions (\ref{aafact}) or (\ref{bbfact}) (it depends on
the fact if the resonance is under the threshold $t_{in}$ or above
it) are more sophisticated analogues of the Breit-Wigner form and
they can be applied to determine resonance parameters $m_r$,
$\Gamma_r$ from experimental data on pseudoscalar meson \emg~FF's
in the region of the resonance under consideration.

As a result, the \emg~FF of any pseudoscalar meson can be represented by a sum
of $i$-terms (\ref{aafact}) of resonances below the threshold $t_{in}$
and $j$-terms (\ref{bbfact}) of resonances above the threshold $t_{in}$
as follows
\begin{eqnarray}
F_P[W(t)] &=& \left(\frac{1-W^2}{1-W^2_N}\right)^2\cdot\label{ffgen}\\
& &\cdot\left\{ \sum_i\frac{(W_N-W_i)(W_N-W^*_i)(W_N-1/W_i)(W_N-1/W^*_i)}
{(W-W_i)(W-W^*_i)(W-1/W_i)(W-1/W^*_i)}(f_{iPP}/f_i) + \right. \nonumber \\
& &\left. + \sum_j\frac{(W_N-W_j)(W_N-W^*_j)(W_N+W_j)(W_N+W^*_j)}
{(W-W_j)(W-W^*_j)(W+W_j)(W+W^*_j)}(f_{jPP}/f_j) \right\},\nonumber
\end{eqnarray}
which takes into account just $n=i+j$ resonances. It is analytic in the
whole complex $s$-plane except for two cuts on the positive real axis
and exhibits the asymptotic behaviour $\sim 1/t$.

If \emg~FF's of hadrons (like baryons and light nuclei) with more
then $n_q=2$ constituent quarks are considered and as a result FF's
exhibit steeper falling in the asymptotic region, then in the
construction of their unitary and analytic models one has to
proceed as follows~\cite{nc106}.

 First, FF's are written down as a
sum of finite number of pole expressions (\ref{pole}). Then the
latter is by explicit requirements transformed to the form, which
is automatically normalized and exhibits assumed asymptotic
behaviour. And only then it is transformed by an application of the
nonlinear transformation (\ref{texp}) to every pole term of FF to
the unitary and analytic model with correct asymptotic behaviour.
In this case a reduction of number of free parameters is achieved.

We note that knowing \emg~FF's of the hadron, by means of the conserved vector
current (CVC) hypothesis one can predict the behaviour of the weak FF's
of the same hadron.

Because the analytic properties of $D_l(s)$ function are of the same type
as analytic properties of the \emg~and weak FF's, the above-prescribed
procedure of a construction of the unitary and analytic models can be
directly applied also to $1/D_l(s)$ function, which will be dominant
in the partial wave amplitude considered in physical region.

\section{Interesting results obtained by unitary and analytic models}

\subsection{Prediction of P-wave isovector $\pi\pi$ phase shift $\delta^1_1(s)$
and inelasticity $\eta^1_1(s)$ above inelastic threshold from
$e^+e^-\to\pi^+\pi^-$ process.}

The unitary and analytic model of $F^{EM}_\pi[W(s)]$ has one
elastic cut $4m^2_\pi<s<+\infty$ and one effective cut
$s_{in}\approx 1{\rm GeV}^2<s<+\infty$. The elastic unitarity
condition
\begin{equation}
{\rm Im}\; F^{EM}_\pi(s) = F^{EM}_\pi(s)\cdot {A^1_1}^*(s)
\end{equation}
can be utilized for the analytic continuation of $F^{EM}_\pi[W(s)]$
through the elastic cut on the second Riemann sheet and as a result one
obtains the FF on the II. sheet to be expressed by FF on the I. sheet and
the $\pi\pi$ partial wave amplitude $A^1_1(s)$ on the I. sheet as follows
\begin{equation}
\left[F^{EM}_\pi(s)\right]^{\rm II} = \frac{\left[F^{EM}_\pi(s)\right]^{\rm I}}
{1+2i\left[A^1_1(s)\right]^{\rm I}}
\end{equation} from where
\begin{equation}
\left[A^1_1(s)\right]^{\rm I} = \frac{\left[F^{EM}_\pi(s)\right]^{\rm I} -
\left[F^{EM}_\pi(s)\right]^{\rm II}}{2i\left[F^{EM}_\pi(s)\right]^{\rm II}}
\label{A}
\end{equation}
to be valid in the whole complex $s$-plane.
\begin{figure}[th]
\begin{center}
\rotatebox{-0.4}{\epsfig{file=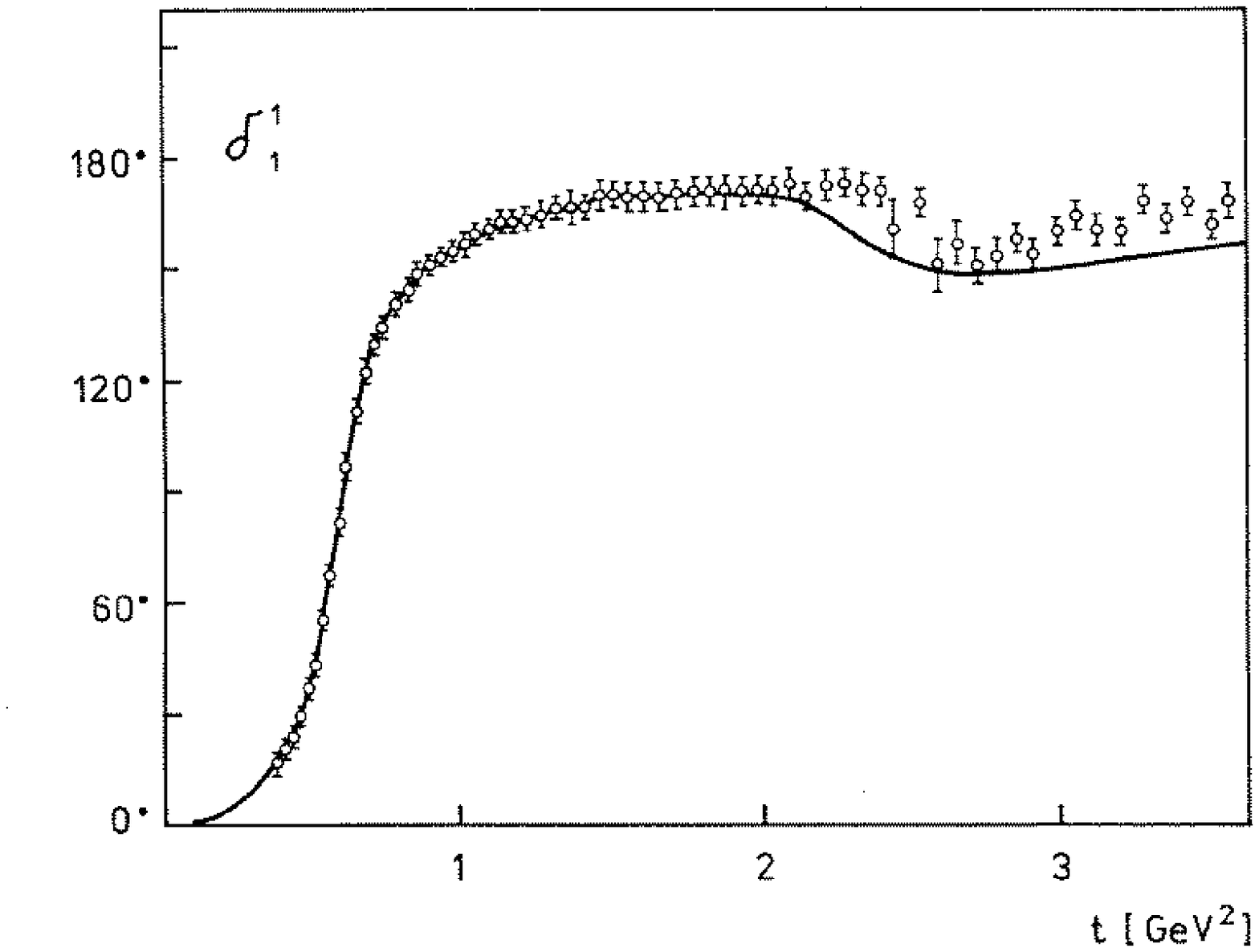,width=0.65\textwidth,height=0.4\textheight}}
\caption{}\label{pi1}
\end{center}
\end{figure}
Now, substituting a standard parameterization of I=J=1 $\pi\pi$ scattering
amplitude at the physical region
\begin{equation}
A^1_1(s+i\varepsilon) = \frac{\eta^1_1(s+i\varepsilon)\;
{\rm e}^{2i\delta^1_1(s+i\varepsilon)}-1}{2i}
\end{equation}
into (\ref{A}) one gets
\begin{equation}
\eta^1_1(s+i\varepsilon){\rm e}^{2i\delta^1_1(s+i\varepsilon)} =
\frac{\left[F^{EM}_\pi[W(s+i\varepsilon)]\right]^{\rm I\hphantom{I}}}{\left[F^{EM}_\pi[W(s+i\varepsilon)]\right]^{\rm II}}
\end{equation}
from where it is straightforward to find
\begin{equation}
\delta^1_1(s+i\varepsilon)=\frac{1}{2}{\rm arctg}
\frac{{\rm Im}\;\frac{\left[F^{EM}_\pi[W(s+i\varepsilon)]\right]^{\rm I\hphantom{I}}}{\left[F^{EM}_\pi[W(s+i\varepsilon)]\right]^{\rm II}}}
{{\rm Re}\;\frac{\left[F^{EM}_\pi[W(s+i\varepsilon)]\right]^{\rm I\hphantom{I}}}{\left[F^{EM}_\pi[W(s+i\varepsilon)]\right]^{\rm II}}}; \;\;\;
\eta^1_1(s+i\varepsilon)=\left|\frac{\left[F^{EM}_\pi[W(s+i\varepsilon)]\right]^{\rm I\hphantom{I}}}{\left[F^{EM}_\pi[W(s+i\varepsilon)]\right]^{\rm II}}\right|.
\label{deltaeta}
\end{equation}
\begin{figure}[t]
\begin{center}
\rotatebox{-0.4}{\epsfig{file=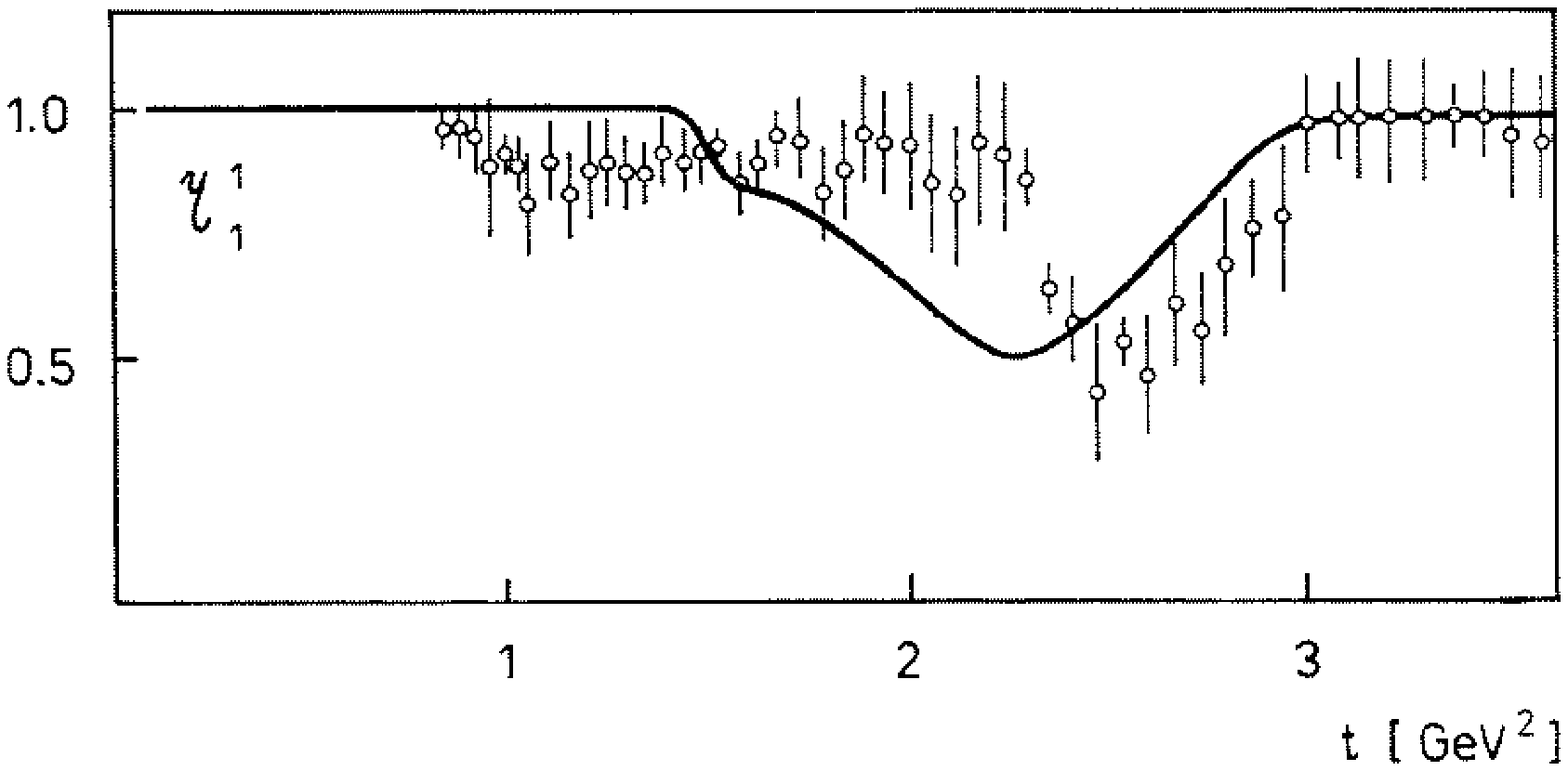,width=0.75\textwidth,height=0.4\textheight}}
\caption{}\label{pi2}
\end{center}
\end{figure}
Substituting unitary and analytic model of the pion \emg~FF in
(\ref{deltaeta}) one predicts~\cite{nc44} behaviour of
$\delta^1_1(s+i\varepsilon)$ and $\eta^1_1(s+i\varepsilon)$ in a
perfect agreement with existing data (see Figs.\ref{pi1} and
\ref{pi2}) also in the inelastic region, i.e. above 1~GeV$^2$.
\begin{figure}[t]
\centerline{\epsfig{file=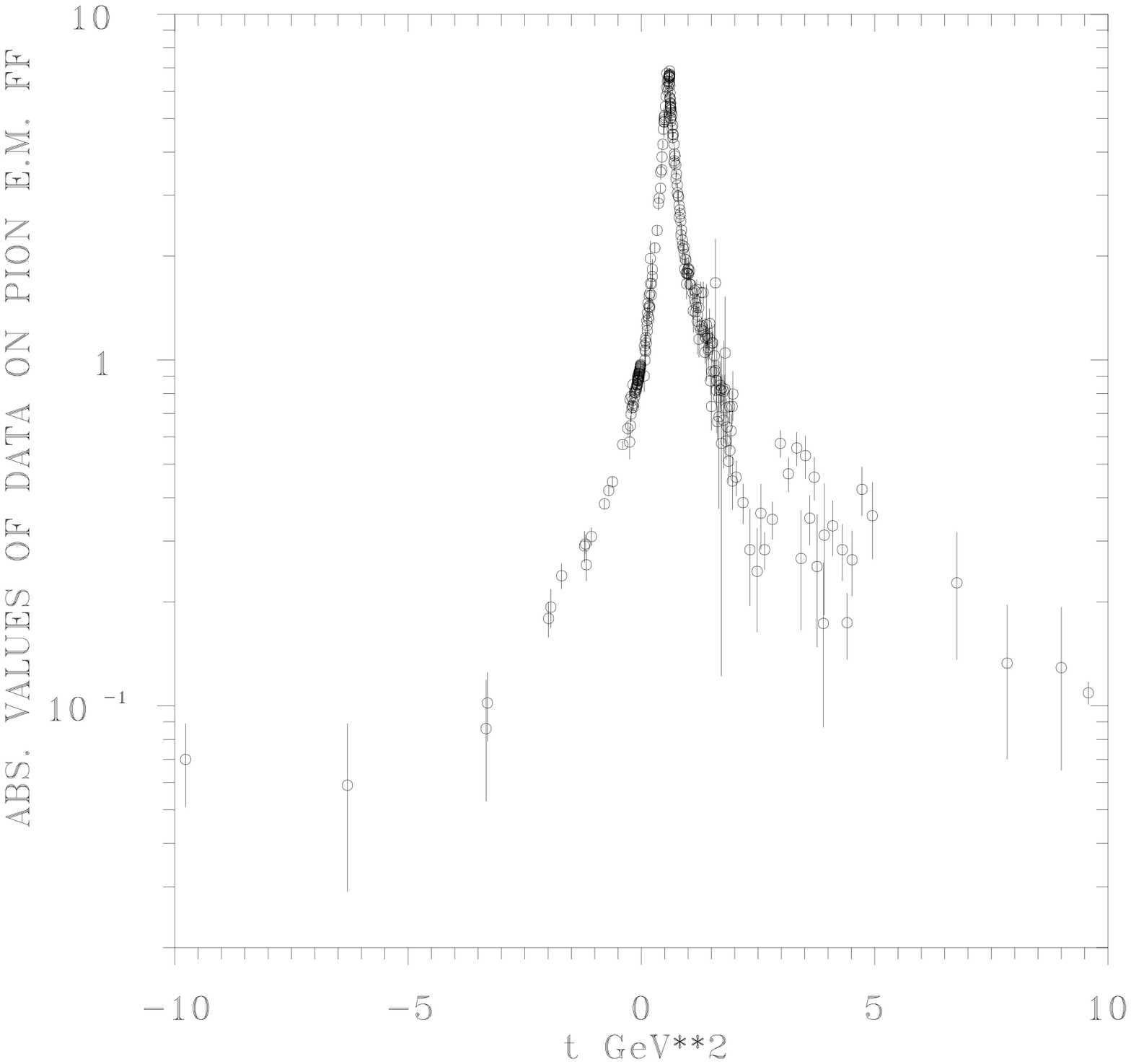,width=0.7\textwidth,
height=0.2\textheight}} \caption{}\label{figrho}
\end{figure}

\subsection{Excited states of the $\rho$(770)-meson}

From Fig.~\ref{figrho}, where the data on the pion electromagnetic
FF are collected, one can see dominating role of the
$\rho$(770)-meson. Besides the latter one can notice in the data
also another resonance around the energy $s=2.9$~GeV$^2$. But it
does not mean at all there are no more other hidden
$\rho$-resonances in the pion electromagnetic FF and the correct
approach in an investigation of the latter problem is to take the
expression (\ref{ffgen}), first, with two resonances, then with
three resonances etc. and to look always for minimal value of
$\chi^2$ in experimental data fit.

\begin{figure}[ht]
\centerline{\epsfig{file=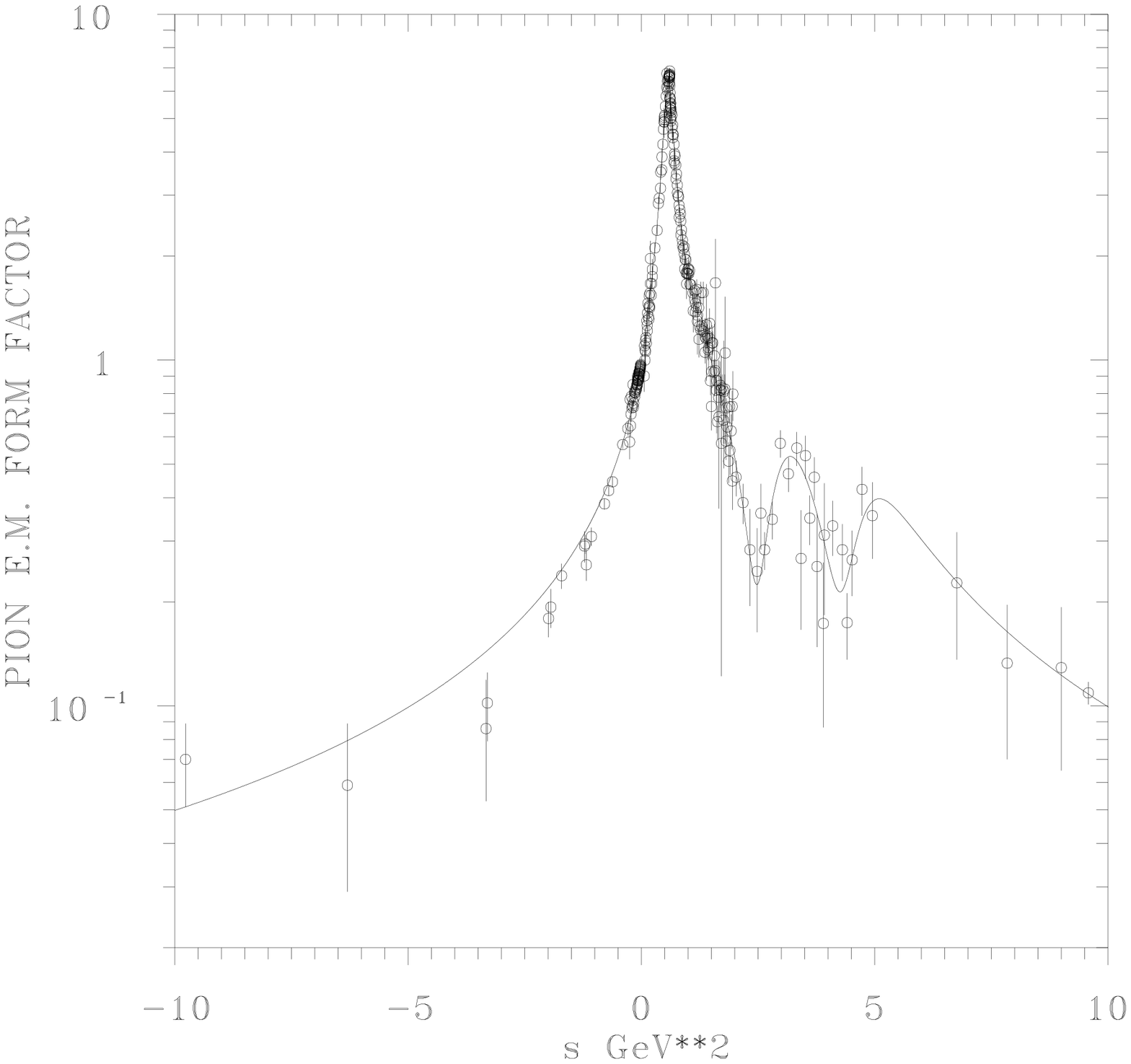,width=0.8\textwidth,
height=0.25\textheight}} \caption{}\label{figrho1}
\end{figure}

Such a program was realized. Considering only two resonances in
(\ref{ffgen}), $\chi^2$/NDF = 539/279 was achieved and $\rho$(770)
with $\rho''$(1700) were identified. However, if three resonances
were taken into account in (\ref{ffgen}), $\chi^2$/NDF~=~382/276
was found and two excited states,  $\rho'$(1450) and
$\rho''$(1700), were detected~\cite{rho1}. Finally, by a
consideration of four-resonances in (\ref{ffgen}), $\chi^2$/NDF =
343/273 was reached and in addition to  $\rho'$(1450) and
$\rho''$(1700), also the third excited state $\rho'''$(2150) of the
$\rho$(770)-meson was revealed~\cite{rho2}(see Fig.~\ref{figrho1}).

\subsection{Confirmation of $\rho'''$(2150) from kaon \emg~form factors data}

\begin{figure}[ht]
\centerline{\epsfig{file=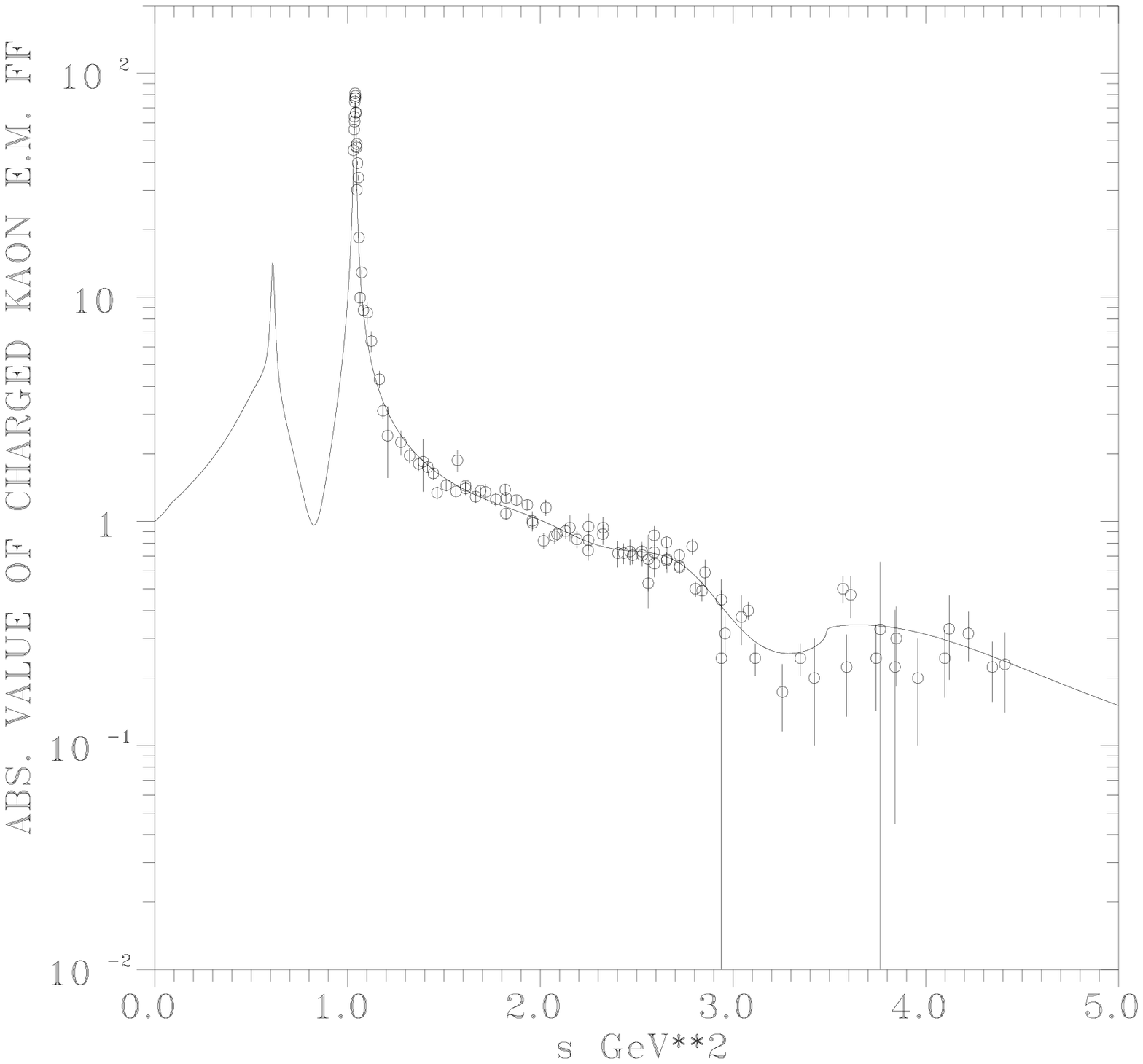,width=0.8\textwidth,
height=0.2\textheight}} \caption{}\label{figkaon1}
\end{figure}

\begin{figure}[ht]
\centerline{\epsfig{file=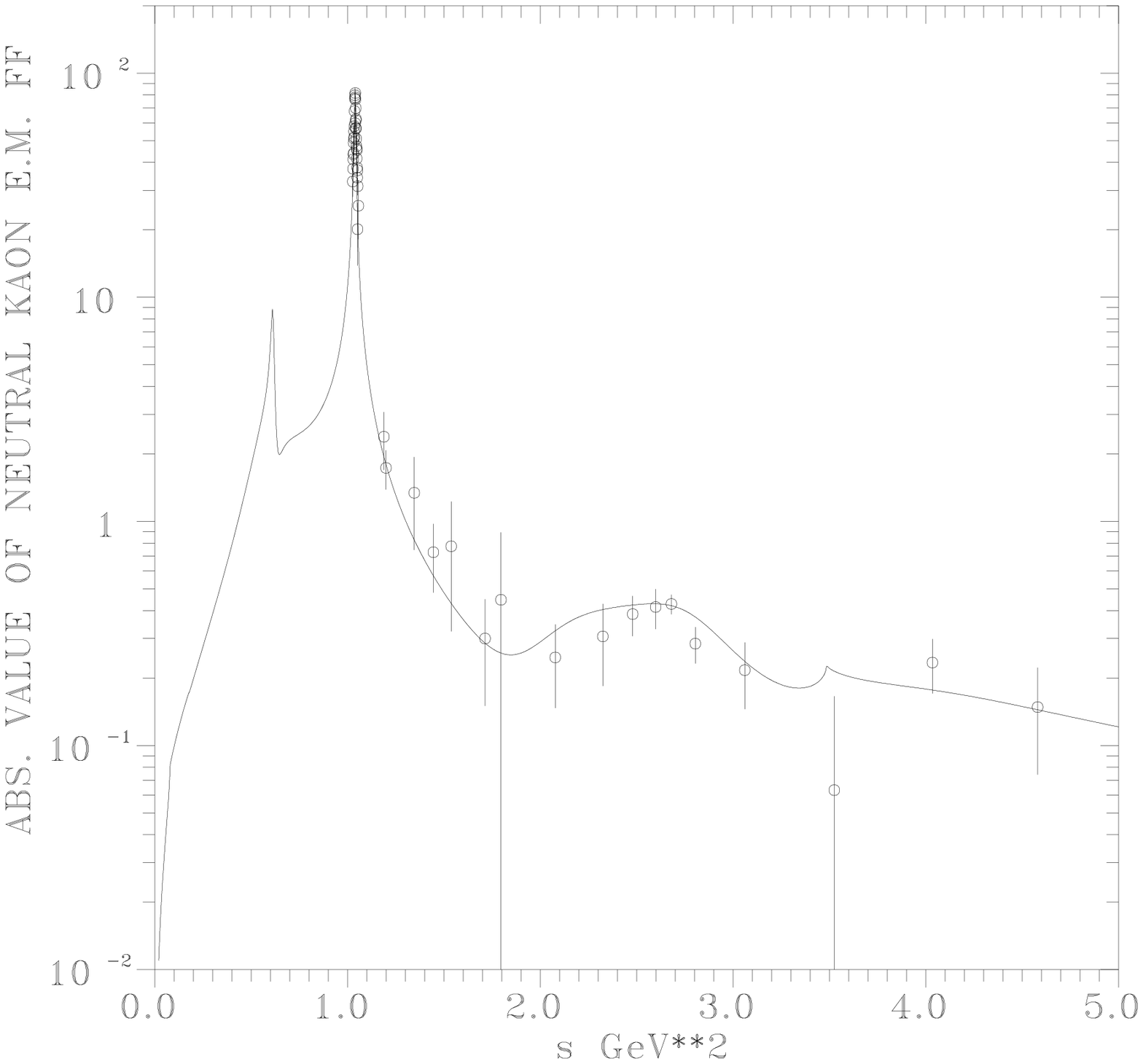,width=0.8\textwidth,
height=0.25\textheight}} \caption{}\label{figkaon2}
\end{figure}

Similarly, the expression (\ref{ffgen}) can be used also for the
isoscalar and isovector parts of the kaon \emg~FF's, to be
saturated by isoscalar and isovector vector mesons, respectively.
In a description of existing data, obtained from $e^+e^-\to K^+K^-$
(Fig.~\ref{figkaon1}) and $e^+e^-\to K^0_SK^0_L$
(Fig.~\ref{figkaon2}) processes, the isovector part of the kaon
\emg~FF's was saturated by three rho-resonances. As a result,
unlike the $e^+e^-\to \pi^+\pi^-$ process, the contribution of the
$\rho'''$(2150) is favored prior to the $\rho''$(1700) by the data
on charge and neutral kaon FF's.

\subsection{Strange-quark vector-current form factor of $K$-mesons}

\begin{figure}[ht]
\centerline{\epsfig{file=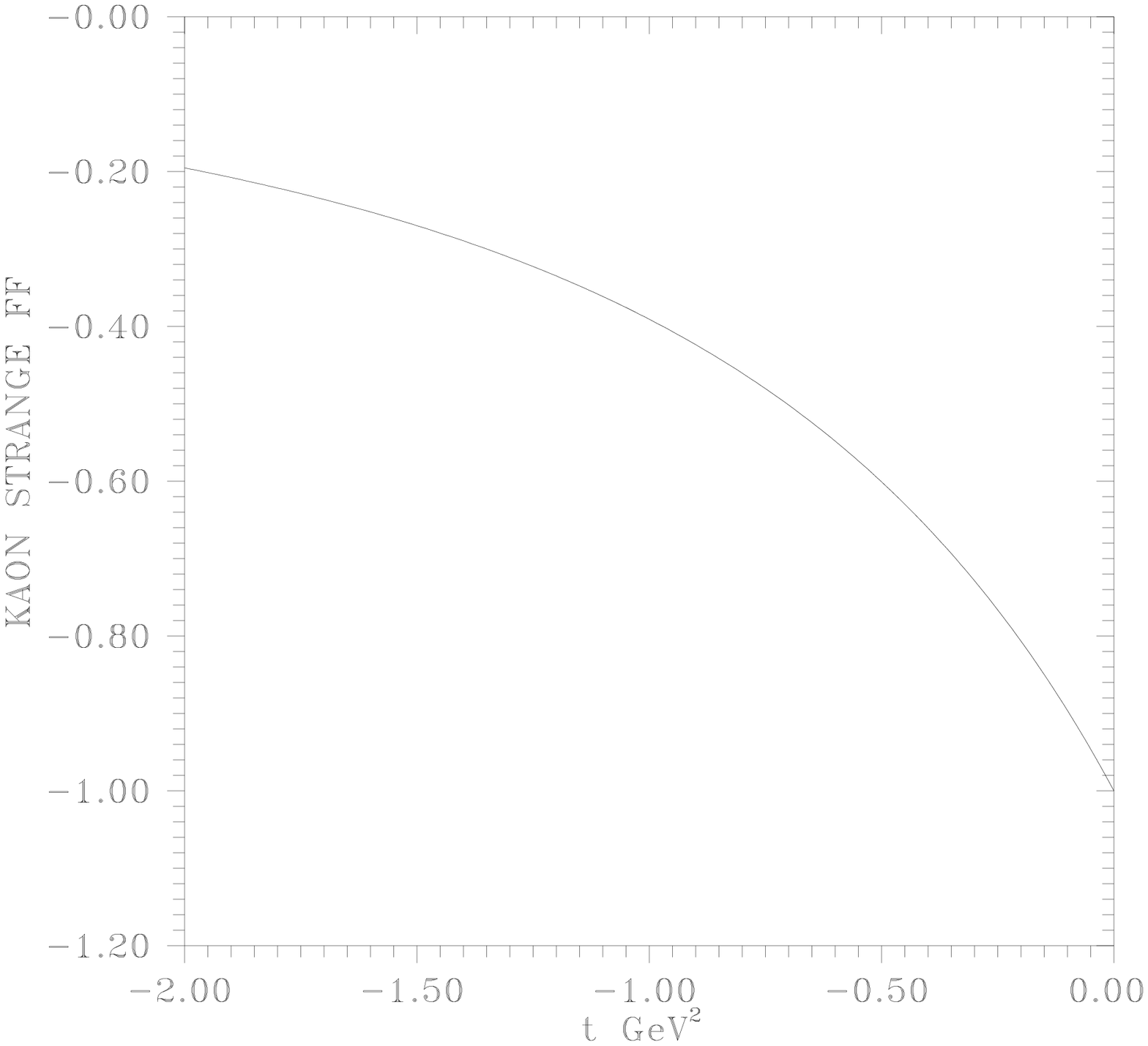,width=0.8\textwidth,
height=0.25\textheight}} \caption{}\label{figstrange1}
\end{figure}

\begin{figure}[hbt]
\centerline{\epsfig{file=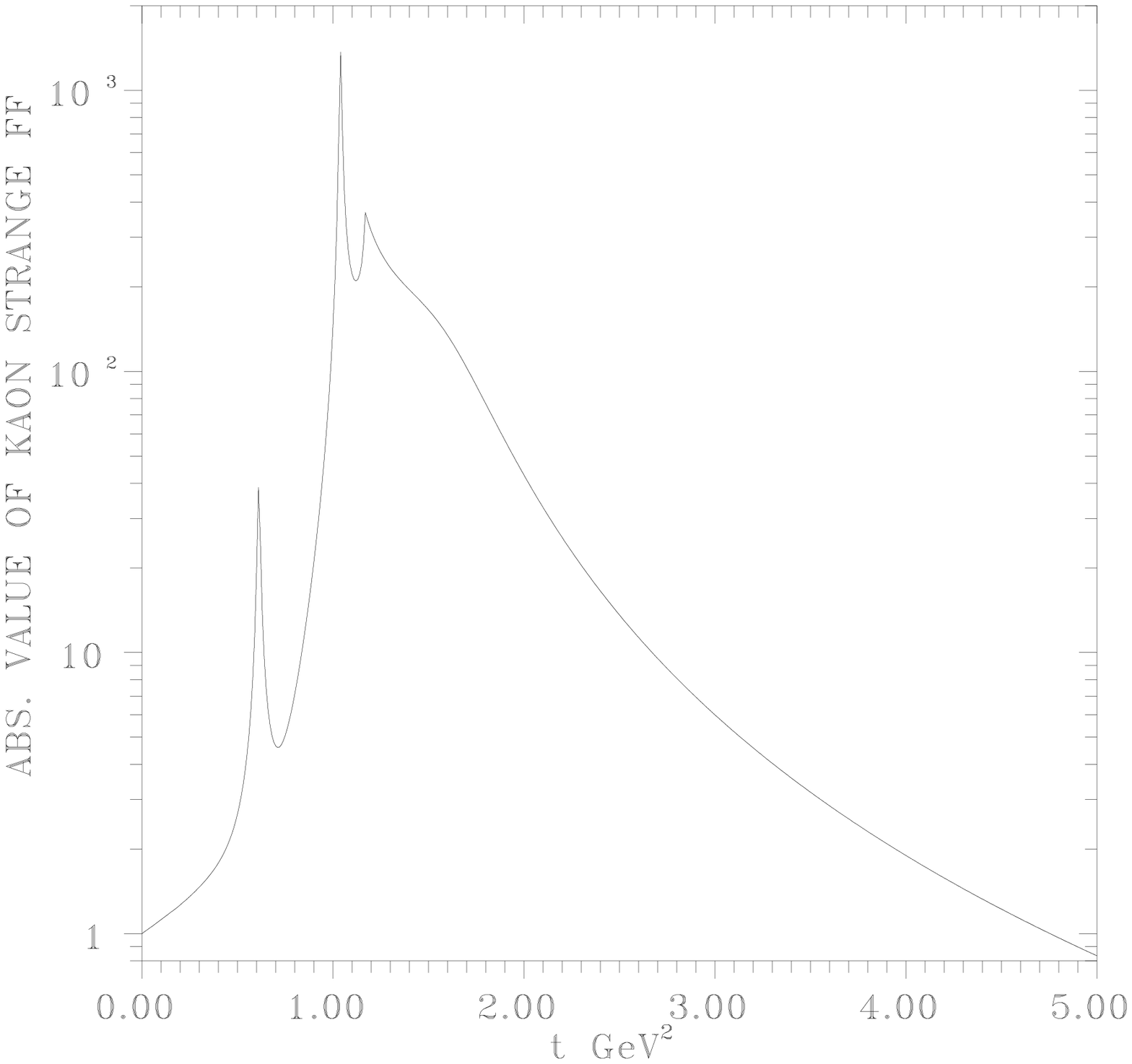,width=0.8\textwidth,
height=0.25\textheight}} \caption{}\label{figstrange2}
\end{figure}

\begin{figure}[hbt]
\centerline{\epsfig{file=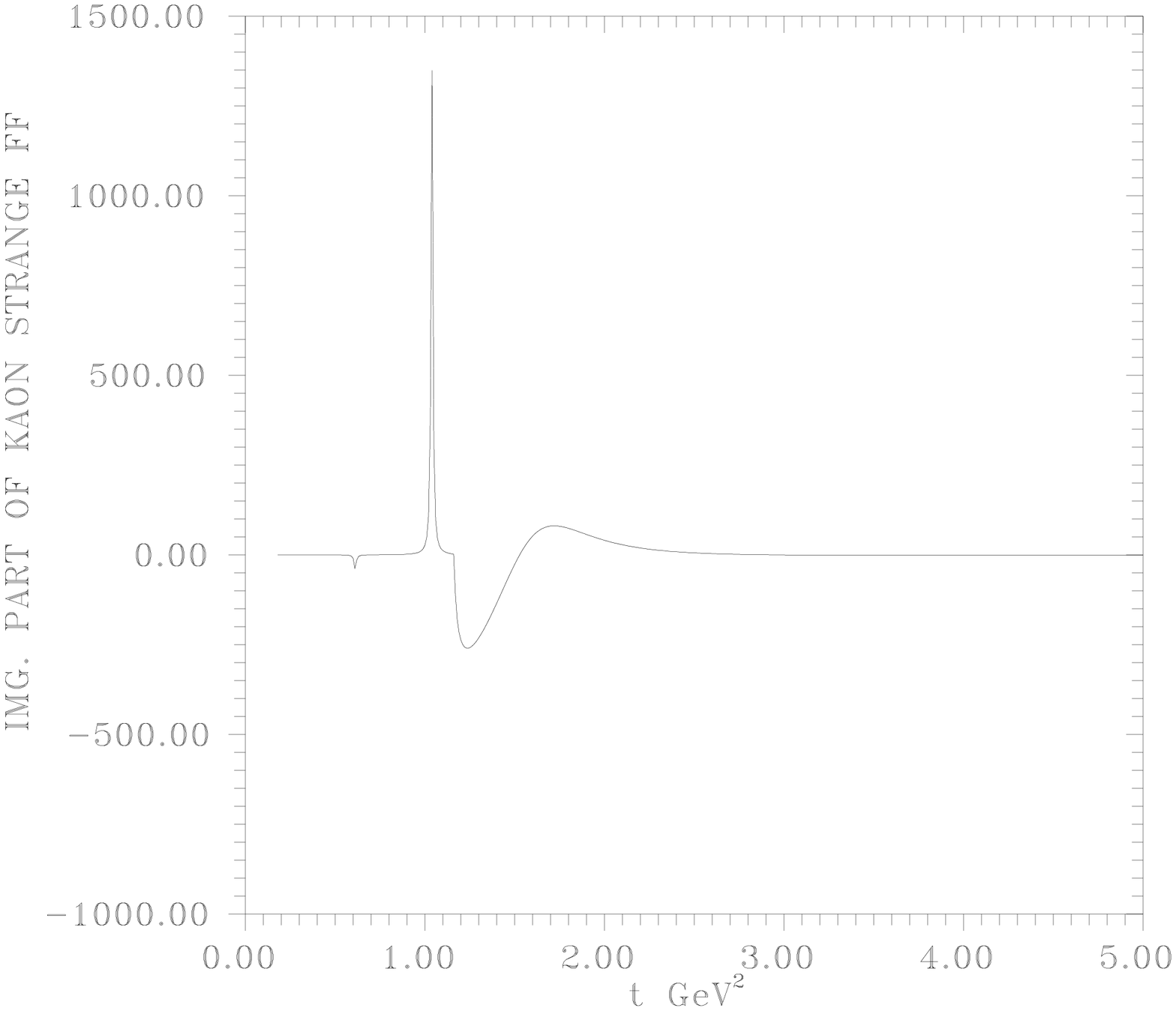,width=0.8\textwidth,
height=0.25\textheight}} \caption{}\label{figstrange3}
\end{figure}

Having unitary and analytic models of the kaon \emg~structure
\begin{eqnarray}
F^{I=0}_K[V(t)]&=& \left(\frac{1-V^2}{1-V^2_N}\right)^2
     \left\{\frac{1}{2}H(V_{\omega'})+\left[L(V_\omega)-H(V_{\omega'})\right](f_{\omega KK}/f^e_\omega) + \right. \nonumber \\
 & &\hphantom{\left(\frac{1-V^2}{1-V^2_N}\right)^2}
\biggl.  + [L(V_\phi) - H(V_{\omega'})](f_{\phi KK}/f^e_\phi)\biggr\} \label{ka1}\\
F^{I=1}_K[W(t)]&=& \left(\frac{1-W^2}{1-W^2_N}\right)^2
     \left\{\frac{1}{2}H(W_{\rho'''})+[L(W_\rho)-H(W_{\rho'''})](f_{\rho KK}/f^e_\rho) + \right. \nonumber \\
 & &\hphantom{\left(\frac{1-W^2}{1-W^2_N}\right)^2}
\biggl.  + [L(W_{\rho'}) - H(W_{\rho'''})](f_{\rho' KK}/f^e_{\rho'})\biggr\} \label{ka2}
\end{eqnarray}
and the strange FF of kaon with the inner structure of $F^{I=0}_K[V(t)]$
\begin{eqnarray}
F^{S}_K[V(t)]&=& \left(\frac{1-V^2}{1-V^2_N}\right)^6
     \left\{-H(V_{\omega'})+[L(V_\omega)-H(V_{\omega'})](f_{\omega KK}/f^s_\omega) + \right. \nonumber \\
 & &\hphantom{\left(\frac{1-V^2}{1-V^2_N}\right)^2}
\biggl.  + [L(V_\phi) - H(V_{\omega'})](f_{\phi KK}/f^s_\phi)\biggr\}
\label{ffstrange}
\end{eqnarray}
set up, one can predict~\cite{strange} a behaviour (see
Figs.~\ref{figstrange1}~-~\ref{figstrange3}) of the strange FF of
$K$-mesons~(\ref{ffstrange}) by means of the Jaffe's
idea~\cite{jaffe} of an evaluation of $(f_{\omega KK}/f^s_\omega)$,
$(f_{\phi KK}/f^s_\phi)$ from $(f_{\omega KK}/f^e_\omega)$,
$(f_{\phi KK}/f^e_\phi)$ determined in a comparison of (\ref{ka1})
and (\ref{ka2}) with data on charge and neutral kaon \emg~FF's.

\subsection{Predictions for weak $\bar{\nu_e}e^-\to P^0P^-$ and
$\tau^-\to\nu_\tau P^0P^-$ processes}

 The corresponding total cross-section (\ref{weakcc}) and the partial decay width
expression (\ref{weakpdw}) of the weak processes $\bar{\nu_e}e^-\to P^0P^-$ and
$\tau^-\to\nu_\tau P^0P^-$, respectively, are completely determined by the weak FF's $F^W_P(s)$ $(P=\pi,K)$.

In V-A theory of weak interactions the hadronic charged weak current takes the form
\begin{equation}
J^\mu_W = V^\mu + A^\mu.
\end{equation}
On the other hand the \emg~current of hadrons has the form
\begin{equation}
J^\mu_E = J^\mu_3 + J^\mu_s
\label{emgcurr}
\end{equation}
where $J^\mu_3$ is a third component of the isotopic vector current
$\vec{J}(J^\mu_1, J^\mu_2, J^\mu_3)$ and $J^\mu_s$ is an isoscalar
current. Almost 50 years ago a very creative postulation was
introduced~\cite{zeld,feyn}
\begin{equation}
V^\mu = J^\mu_1 - iJ^\mu_2
\end{equation}
i.e. the charged weak vector current $V^\mu$ and $J^\mu_3$ of the
\emg~current (\ref{emgcurr}) are components of the same isotopic
vector $\vec{J}(J^\mu_1, J^\mu_2, J^\mu_3)$. Because strong
interactions are invariant according to the isotopic SU(2) group,
then by $\vec{J}(J^\mu_1, J^\mu_2, J^\mu_3)$ the equation
\begin{equation}
\partial_\mu J^\mu_i = 0
\end{equation}
is fulfilled, from where the conserved-vector-current (CVC) hypothesis
\begin{equation}
\partial_\mu V^\mu = 0 \label{cvc}
\end{equation}
follows.
\begin{figure}[ht]
\centerline{\epsfig{file=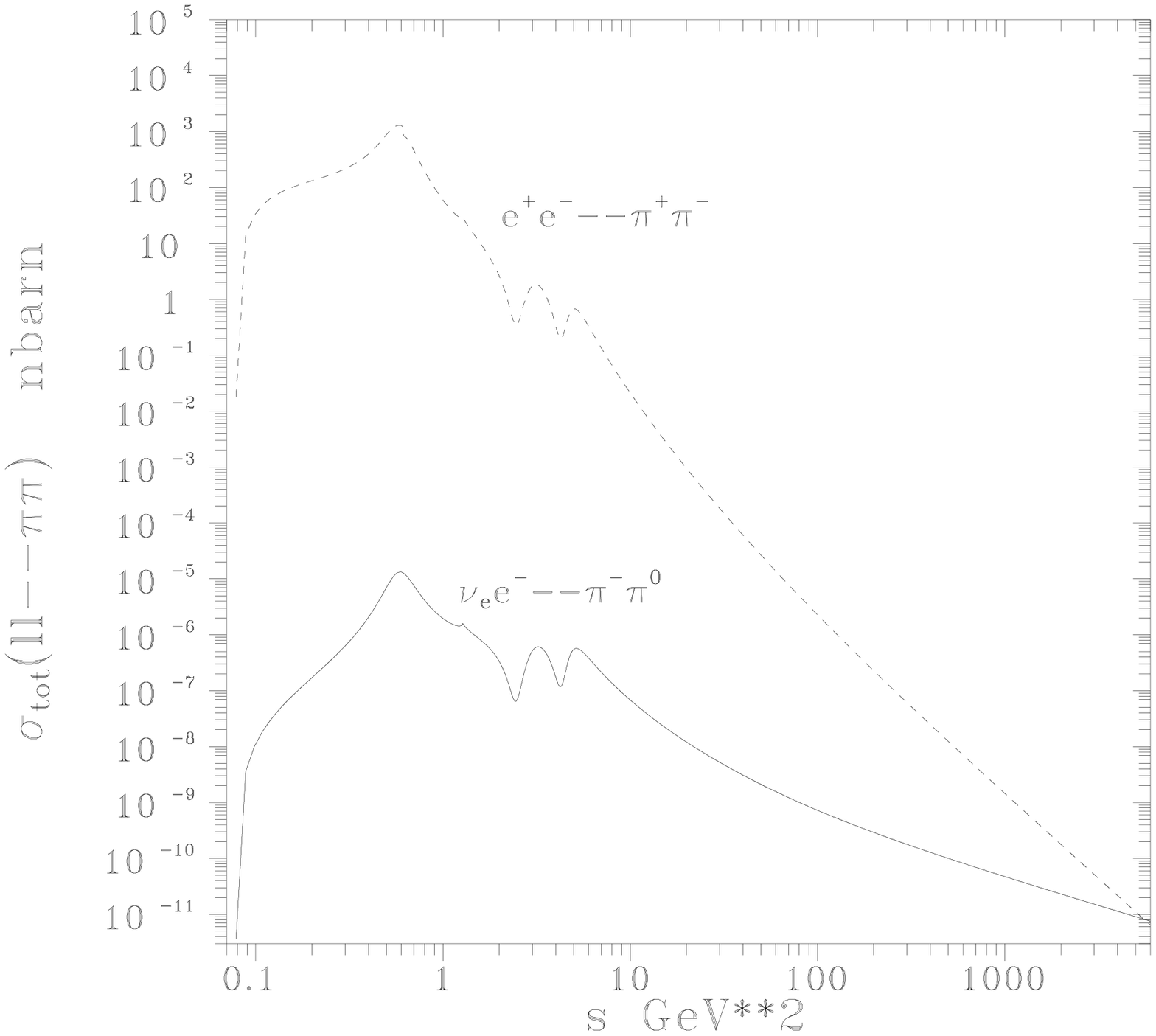,width=0.7\textwidth,height=0.25\textheight}}
\caption{}\label{figstot}
\end{figure}
Starting from eq. (\ref{cvc}) one can prove the relation between
the \emg~and weak pion FF's
\begin{equation}
F^W_\pi(s) = \sqrt{2}F^{E,I=1}_\pi(s) \label{weakpirel}
\end{equation}
and also the relation between weak kaon FF and the isovector part of the
\emg~FF's of $K$-mesons
\begin{equation}
F^W_K(s) = 2 F^{E,I=1}_K(s).\label{weakkrel}
\end{equation}
If now for $F^{E,I=1}_\pi(s)$ and $F^{E,I=1}_K(s)$ unitary and analytic models of the
type (\ref{ffgen}) are taken into account, by means of (\ref{weakpirel}) and
(\ref{weakkrel}) behaviours of $F^W_\pi(s)$ and $F^W_K(s)$ are found and
evaluations of physical quantities (\ref{weakcc}) and (\ref{weakpdw}) are achieved.\\[3mm]
\begin{tabular}{ll}
\parbox{0.4\textwidth}{Our theoretical predictions for branching ratios:}~~~~ &
\parbox{0.4\textwidth}{Exp. values of branching ratios from ALEPH in CERN:}\\
$B(\tau^-\to\nu_\tau\pi^0\pi^-)$ = 25.76\%~\cite{brpi}
&
$B(\tau^-\to\nu_\tau\pi^0\pi^-)$ = 25.34 $\pm$ 0.19\%~\cite{aleph} \\
$B(\tau^-\to\nu_\tau K^0K^-)$ = 0.18\%~\cite{brk}
&
$B(\tau^-\to\nu_\tau K^0K^-)$ = 0.194 $\pm$ 0.042\%.~\cite{aleph}
\end{tabular}\\[3mm]
For comparison of $\sigma_{\rm tot}(\bar{\nu_e}e^-\to\pi^0\pi^-)$ and
$\sigma_{\rm tot}(e^+e^-\to\pi^+\pi^-)$ see Fig.~\ref{figstot}. For kaons can't be carried out such a comparison.

\subsection{Unitary and analytic model of nucleon \emg~structure and its predictability}

\begin{figure}[hbt]
\centerline{\epsfig{file=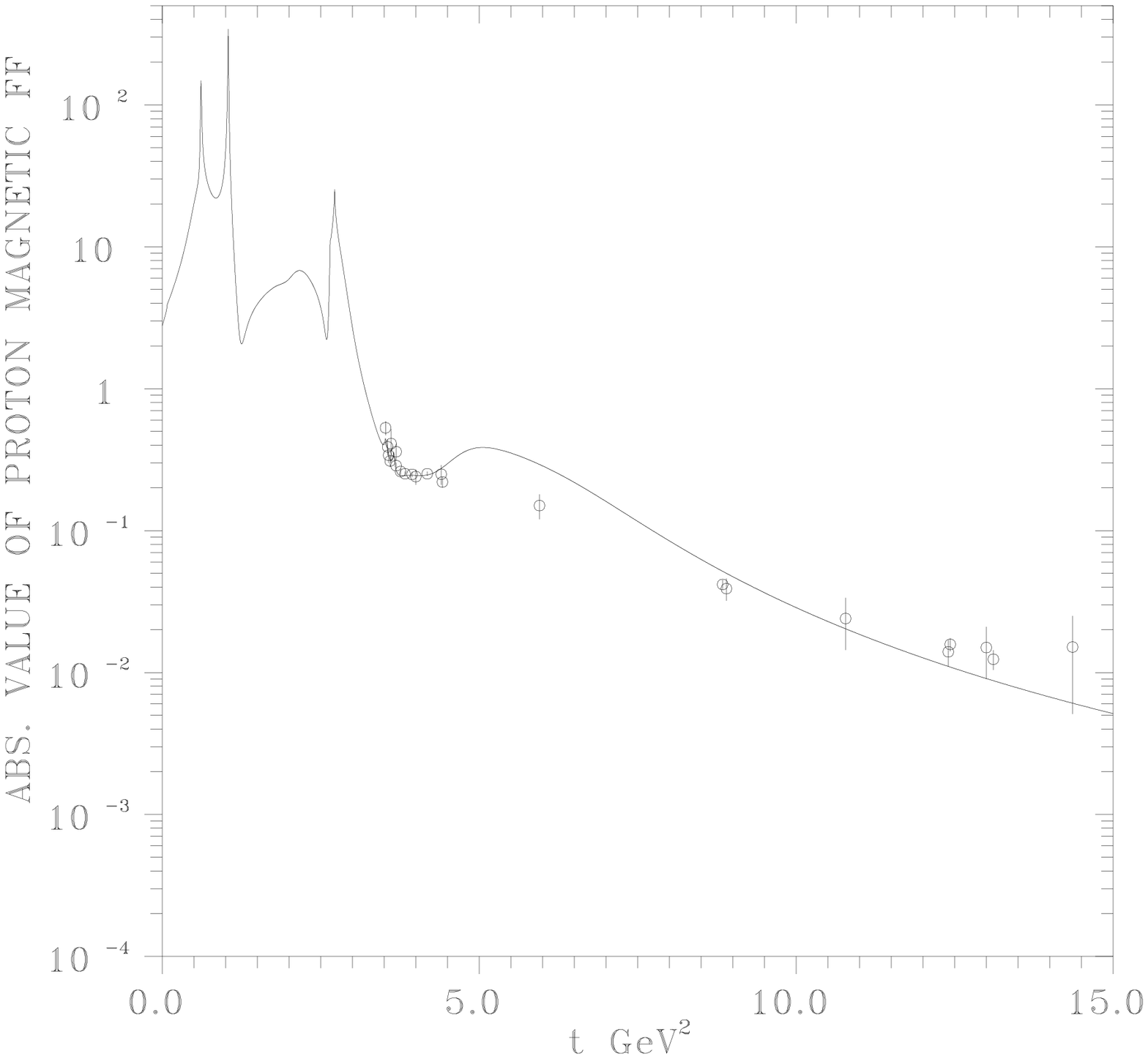,width=0.7\textwidth,
height=0.25\textheight}} \caption{}\label{fig10resprot}
\end{figure}

\begin{figure}[hbt]
\centerline{\epsfig{file=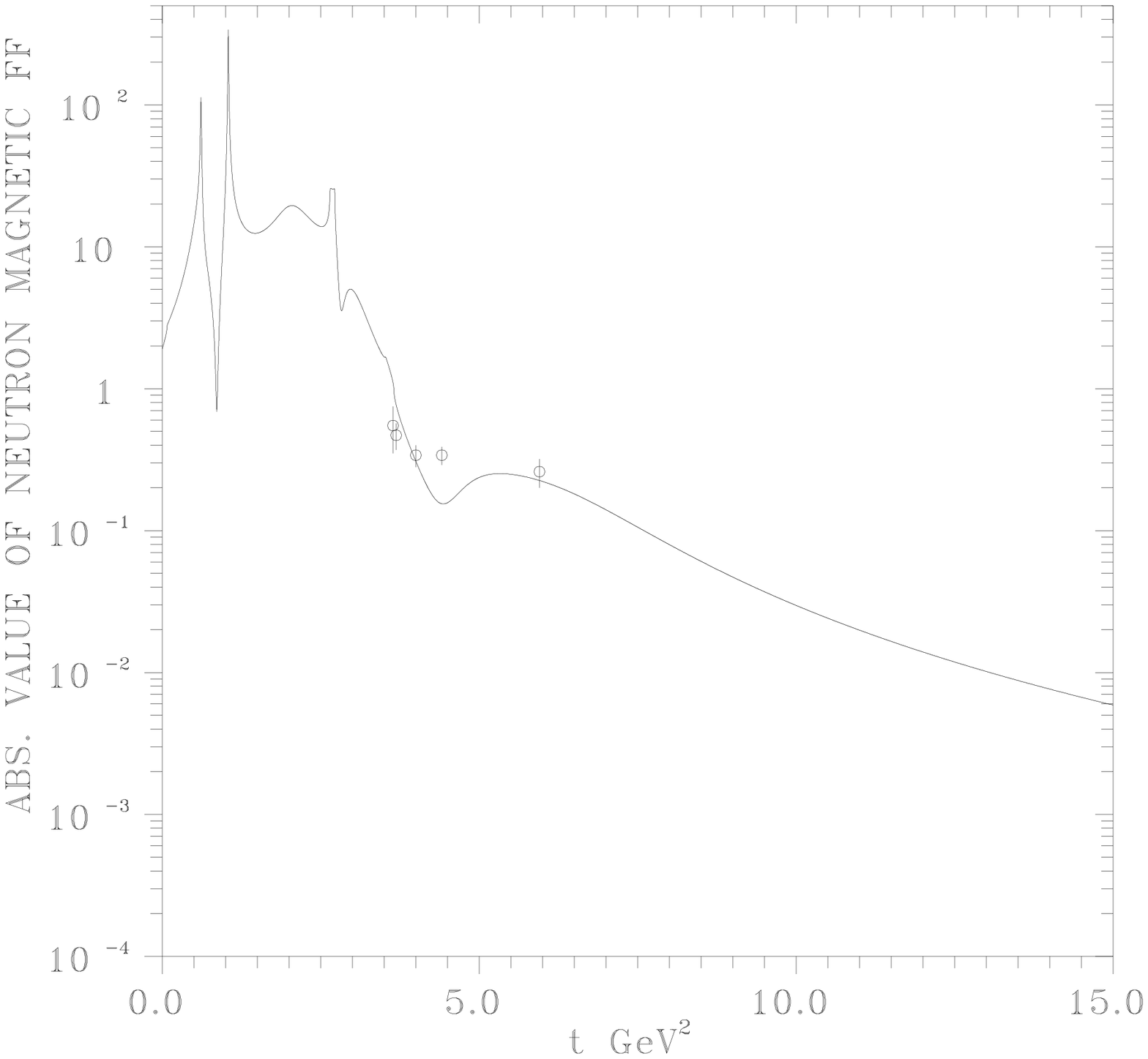,width=0.7\textwidth,
height=0.25\textheight}} \caption{}\label{fig10resneut}
\end{figure}

\begin{figure}[hbt]
\centerline{\epsfig{file=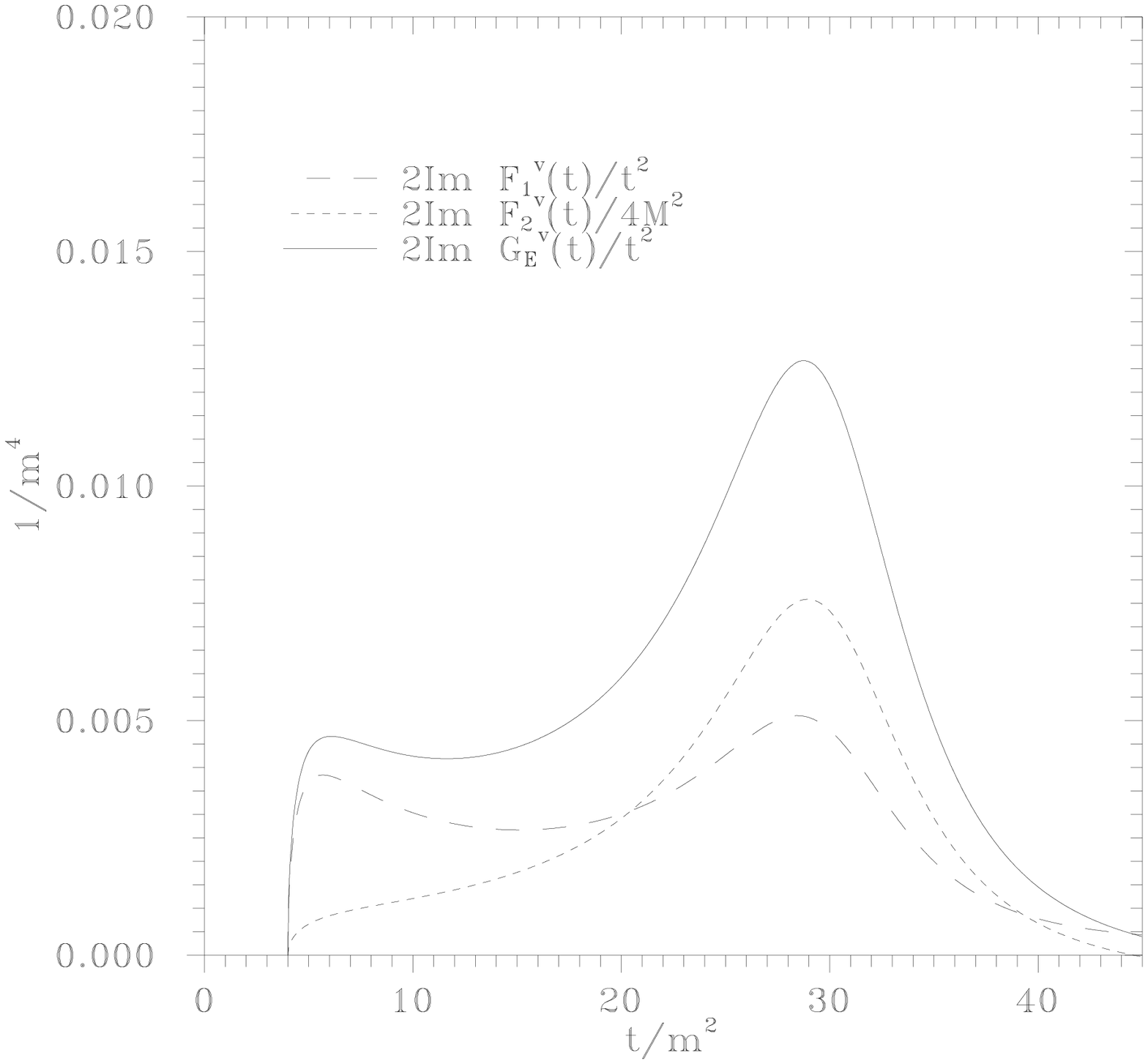,width=0.7\textwidth,
height=0.25\textheight}} \caption{}\label{figspect1}
\end{figure}

\begin{figure}[hbt]
\centerline{\epsfig{file=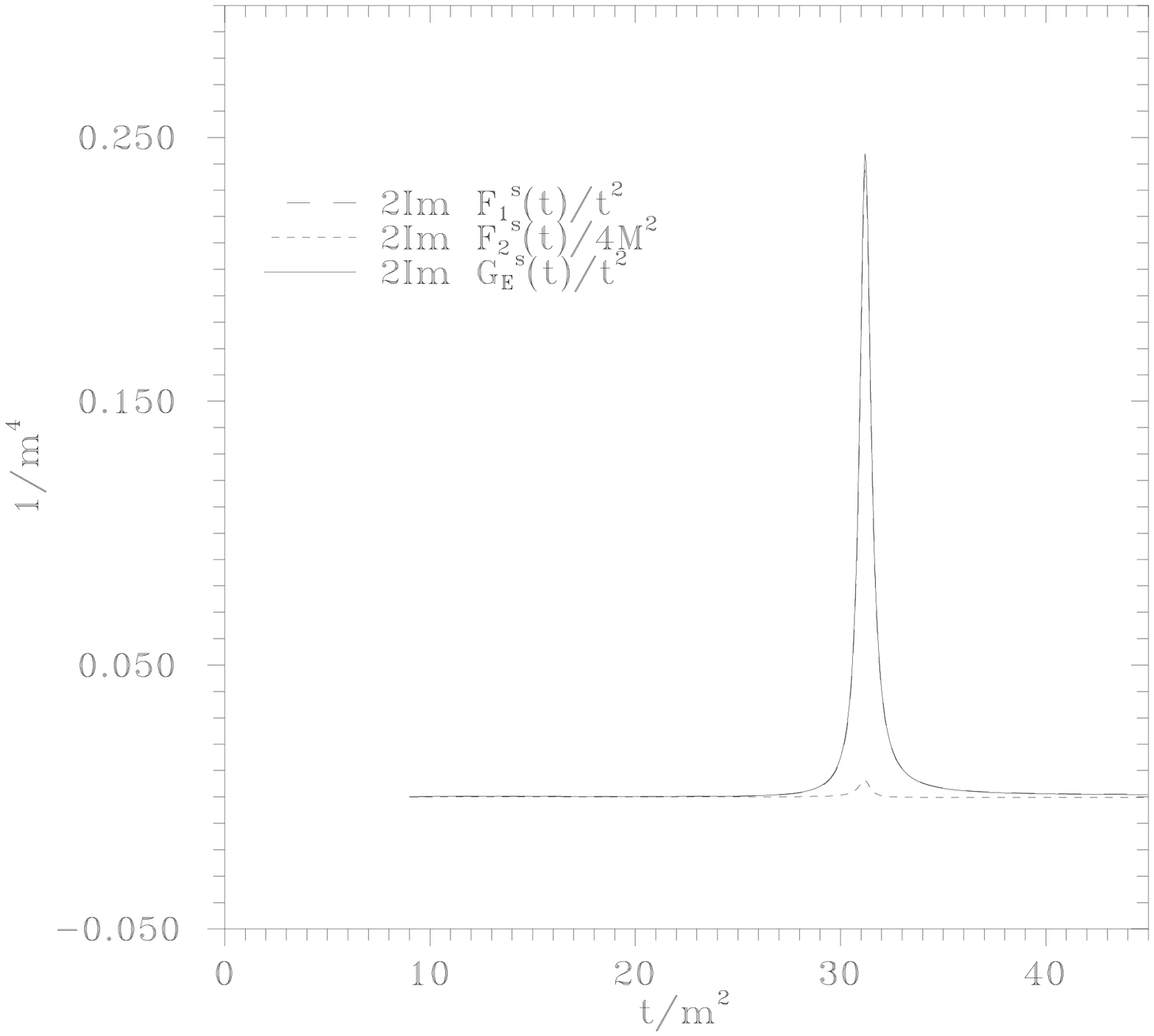,width=0.7\textwidth,
height=0.25\textheight}} \caption{}\label{figspect2}
\end{figure}

Keeping a procedure discussed at the end of the
Section~\ref{secmainidea}. one can construct ten-resonance (5
isoscalars and 5 isovectors) unitary and analytic model of nucleon
\emg~structure~\cite{10res}, which describes well all existing
experimental space-like and time-like data on the nucleon \emg~form
factors, including also FERMILAB proton eight points~\cite{fermi1,
fermi2} at higher energies (see Fig.~\ref{fig10resprot}) and FENICE
(Frascati) results~\cite{fenice} on the neutron (see
Fig.~\ref{fig10resneut}), for the first time. The latter was
achieved without any external constraints on the neutron charge
radius and on the isovector spectral functions following from the
$\pi N$-scattering data and pion \emg~form factor behaviour through
the unitarity condition~\cite{unit}. Just opposite,  the model
itself spontaneously predicts a negative value for the neutron
charge radius and a pronounced effect of the two-pion continuum on
the isovector spectral functions (see Fig.~\ref{figspect1})
revealing the strong enhancement of the left wing of the
$\rho$(770)-resonance close to two-pion threshold.
Moreover, an existence of the fourth excited state of the
$\rho$(770) meson at $t\approx 6.25$~GeV$^2$, the large values of
coupling constants $f^{(1)}_{\phi NN}=12.1$ and $f^{(2)}_{\phi
NN}=3.4$ (indicating a violation of the OZI rule and a nonzero
strange-quark content in nucleons) as well as the isoscalar
spectral functions behaviours (see Fig.~\ref{figspect2}) are
predicted by such a model.

\subsection{Confirmation of strange-quark contributions to nucleon structure by
the unitary and analytic model of nucleon \emg~FF's}

More recently, a well-defined experimental program has just started
determination of the nucleon matrix element
\begin{equation}
<p'|\bar{s}\gamma_\mu s|p> = \bar{u}(p')
\left[\gamma_\mu F^s_1(t) + i\frac{\sigma_{\mu\nu}q^\nu}{2\;m_N}F^s_2(t)\right]
u(p)
\end{equation}
of the strange-quark vector current $\bar{s}\gamma_\mu s$ by means
of the elastic scattering of polarized electron beam on liquid
hydrogen target, in which (unlike the most theoretical estimates)
unexpected positive values of strange FF's
\begin{eqnarray}
G^s_E(t) &=& F^{(s)}_1(t) + \frac{t}{4m^2_N}F^{(s)}_2(t) \nonumber \\
G^s_M(t) &=& F^{(s)}_1(t) + F^{(s)}_2(t)
\end{eqnarray}
were revealed:
\begin{itemize}
\item[]
the SAMPLE Collaboration at MIT/Bates Linear Accelerator Center~\cite{MIT1}\\
\begin{equation}
G^s_M(-0.1) = +0.23 \pm 0.37 \pm 0.15 \pm 0.19 \;\;[\mu_N]
\label{exp1}
\end{equation}
\item[]
or improved measurements at the same Center~\cite{MIT2}\\
\begin{equation}
G^s_M(-0.1) = +0.61 \pm 0.17 \pm 0.21 \pm 0.19 \;\;[\mu_N]
\label{exp2}
\end{equation}
\item[]
and the HAPPEX Collaboration at TJNAF~\cite{happex}
\begin{equation}
G^s_E(-0.48)+0.39 G^s_M(-0.48) = +0.023 \pm 0.034 \pm 0.022 \pm 0.026.
\label{exp3}
\end{equation}
\end{itemize}

 If there are really nonzero strange nucleon FF's, then as a result of the isospin
zero value of the strange quark, they can exclusively contribute
just to isoscalar parts of the nucleon \emg~FF's and never to the
isovector FF's. We show, that by specific eight resonance unitary
and analytic model of the nucleon \emg~structure, the isoscalar
part of which depends only on $\omega$ and $\phi$ coupling constant
ratios as free parameters, one can explain all experimental results
(\ref{exp1})-(\ref{exp3}).
\begin{figure}[hbt]
\centerline{\epsfig{file=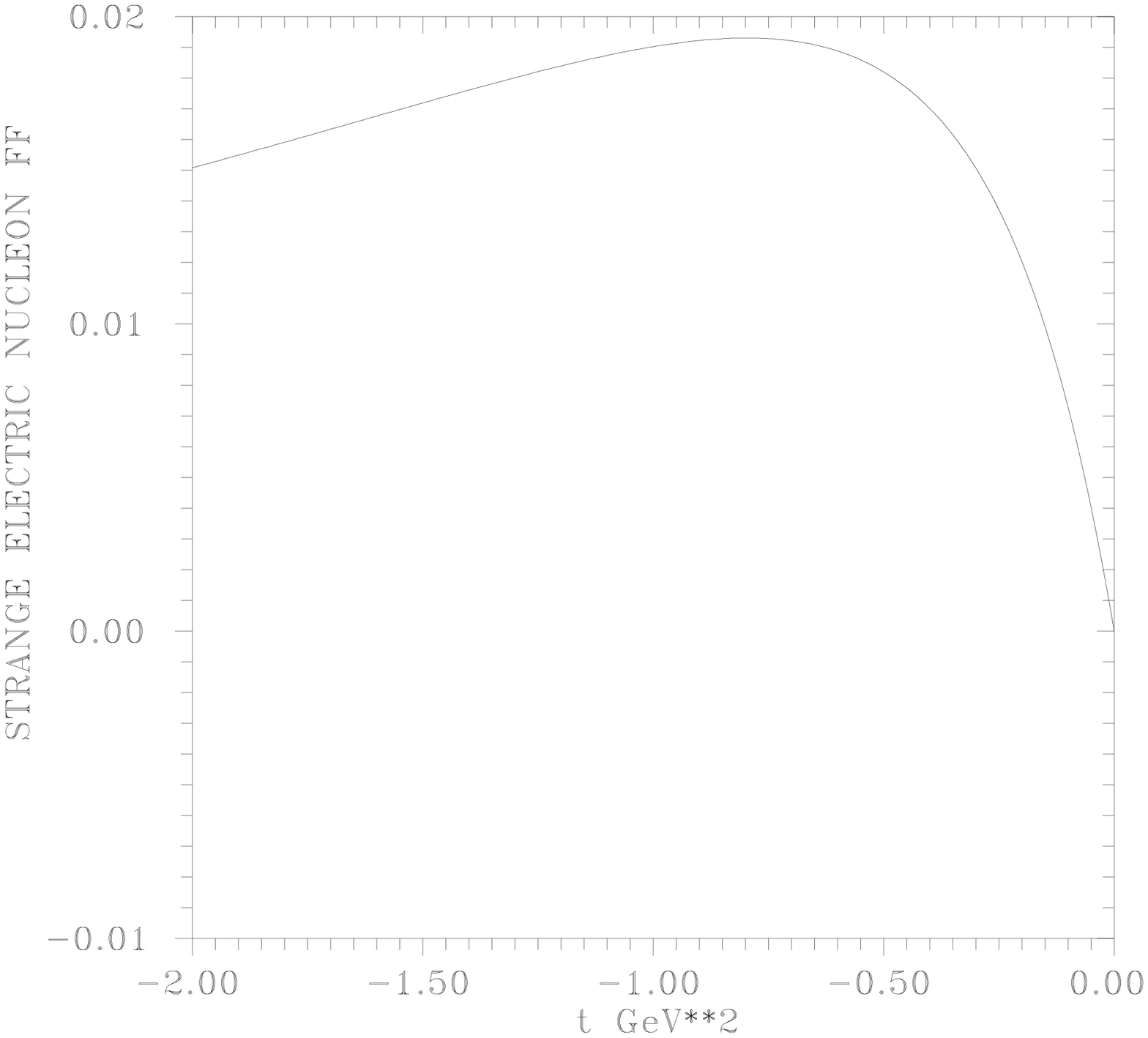,width=0.7\textwidth,
height=0.25\textheight}} \caption{}\label{fig8rese}
\end{figure}

The main idea of a prediction of strange nucleon FF behaviour, based on the
$\omega-\phi$ mixing and on the assumption that the quark current of some flavour
couples with universal strength exclusively to the vector-meson wave function
component of the same flavour, consists in the following. Provided that one knows
free parameters $(f^{(i)}_{\omega NN}/f^e_\omega)$,
$(f^{(i)}_{\phi NN}/f^e_\phi)$ of the unitary and analytic model
\begin{equation}
F^{I=0}_i(t) = f[t;(f^{(i)}_{\omega NN}/f^e_\omega), (f^{(i)}_{\phi NN}/f^e_\phi)]
\;\;\;\;(i=1,2)
\label{8res}
\end{equation}
then the unknown free parameters $(f^{(i)}_{\omega
NN}/f^s_\omega)$, $(f^{(i)}_{\phi NN}/f^s_\phi)$ of a strange
nucleon FF's unitary and analytic model
\begin{equation}
F^s_i(t) = \bar{f}[t;(f^{(i)}_{\omega NN}/f^s_\omega), (f^{(i)}_{\phi NN}/f^s_\phi)]
\;\;\;\;(i=1,2)
\end{equation}
of the same inner analytic structure as $F^{I=0}_i(t)$, but of course with
different norm and possibly with different asymptotics (therefore denoted by
$\bar{f}$), are numerically evaluated by the relations~\cite{jaffe}
\begin{eqnarray}
(f^{(i)}_{\omega NN}/f^s_\omega) &=&
-\sqrt{6}\;\frac{\sin\varepsilon}{\sin(\varepsilon+\theta_0)}
(f^{(i)}_{\omega NN}/f^e_\omega) \nonumber \\
& &\;\;\;\;\;\;\;\;\;\;\;\;\;\;\;\;\;\;\;\;\;\;\;\;\;\;\;\;\;\;\;\;\;\;\;\;\;\;\;
\;\;\;\;\;\;\;\;\;\;\;\;\;\; (i=1,2) \label{jaffe1}\\
(f^{(i)}_{\phi NN}/f^s_\phi) &=&
-\sqrt{6}\;\frac{\cos\varepsilon}{\cos(\varepsilon+\theta_0)}
(f^{(i)}_{\phi NN}/f^e_\phi) \nonumber
\end{eqnarray}
where $\varepsilon=3.7^\circ$ is a deviation from the ideally mixing angle
$\theta_0=35.3^\circ$.

\begin{figure}[hbt]
\centerline{\epsfig{file=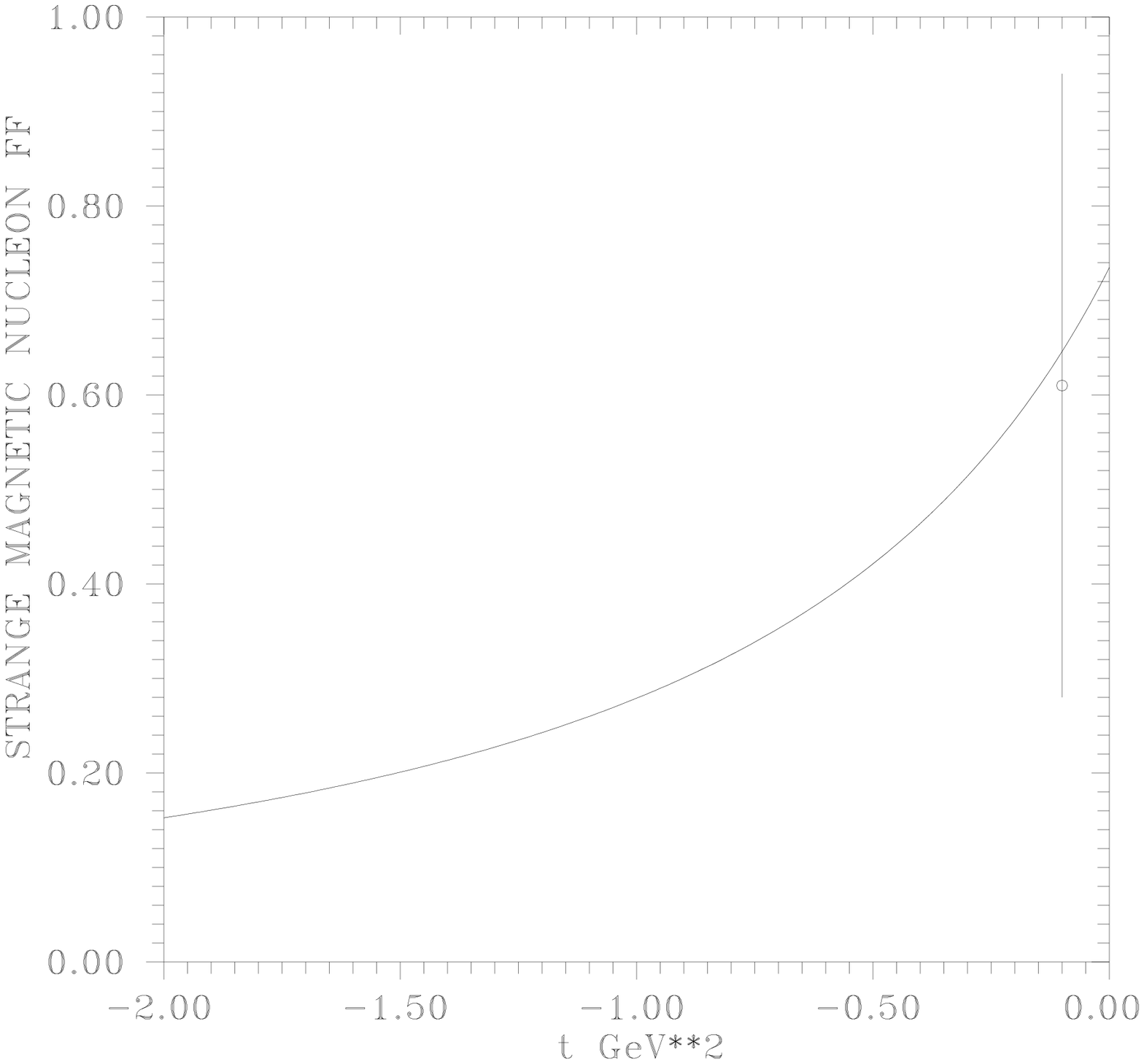,width=0.7\textwidth,
height=0.25\textheight}} \caption{}\label{fig8resm}
\end{figure}

The parameters $(f^{(i)}_{\omega NN}/f^e_\omega)$, $(f^{(i)}_{\phi
NN}/f^e_\phi)$ $(i=1,2)$ were evaluated by a comparison of the
eight resonance unitary and analytic model (\ref{8res}) with
existing data, then $(f^{(i)}_{\omega NN}/f^s_\omega)$,
$(f^{(i)}_{\phi NN}/f^s_\phi)$ were calculated by means of
(\ref{jaffe1}) and behaviour of $G^s_E(t)$, $G^s_M(t)$ was
predicted~\cite{mod8res} as it is presented in Figs.~\ref{fig8rese}
and \ref{fig8resm}. One can see from Fig.~\ref{fig8resm} that the
improved SAMPLE collaboration result~(\ref{exp2}) is preferred to
the older one~(\ref{exp1}) and that the strange magnetic moment of
the nucleon is positive valued to be
$\mu_s=G^s_M(0)=F^s_2(0)=+0.73\;[\mu_N]$. On the other hand one
obtains
\begin{equation}
G^s_E(-0.48)+0.39\;G^s_M(-0.48) = +0.185
\end{equation}
indicating also the positive value like in~(\ref{exp3}).
 The behaviours of $G^s_E(t)$, $G^s_M(t)$ give also a prediction for a combination
\begin{equation}
G^s_E(-0.23)+0.22\;G^s_M(-0.23) = +0.135
\end{equation}
which is just measured at the MAMI A4~\cite{MAMI} running experiment.

\subsection{Prediction of $\sigma_{\rm tot}(e^+e^-\to Y\bar{Y})$ behaviour}

According to SU(3) classification of hadrons there is ${1/2}^+$
octet of baryons including nucleon doublet $[n,p]$ together with 6
other hyperons $[\Lambda^0]$, $[\Sigma^+,\Sigma^0,\Sigma^-]$ and
$[\Xi^0,\Xi^-]$. Though there is almost zero experimental
information on the hyperon \emg~structure, by using the unitary and
analytic models of \emg~FF's of all members of the ${1/2}^+$ octet
of baryons, the experimental information on the nucleon \emg~FF's
and SU(3) symmetry, one can predict behaviours of the hyperon
\emg~FF's and as a result also behaviours of
\begin{equation}
\sigma_{\rm tot}(e^+e^-\to Y\bar{Y}) = \frac{4\pi\alpha^2\beta_Y}{3t}
\left[|G^Y_M(t)|^2 + \frac{2m^2_Y}{t}|G^Y_E(t)|^2\right].
\label{stothyp}
\end{equation}
\begin{figure}[hbt]
\centerline{\epsfig{file=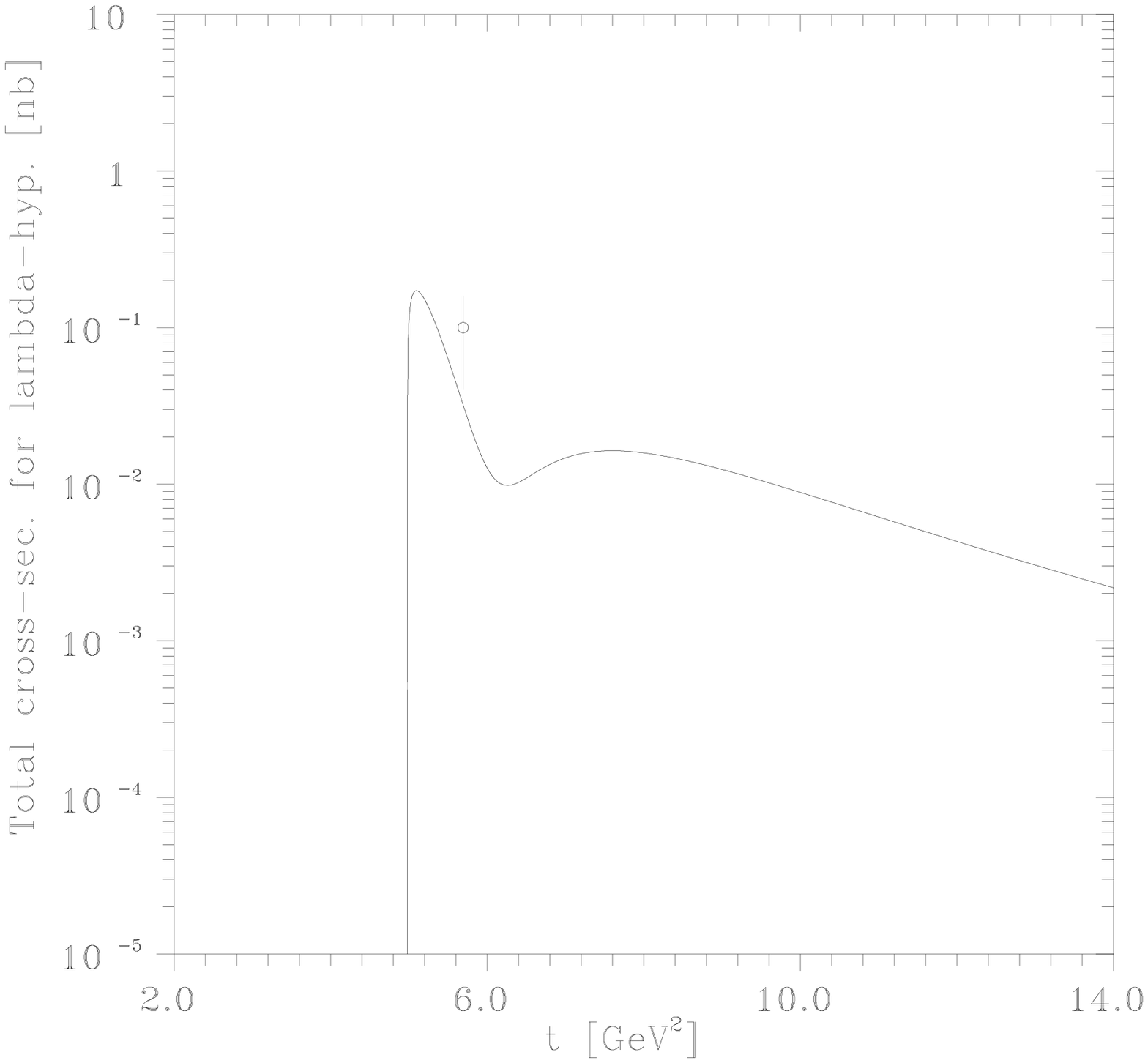,width=0.7\textwidth, height=0.25
\textheight}} \caption{}\label{fighyp1}
\end{figure}

Practically, one has to start with a specific nine resonance unitary and
analytic model of the \emg~FF's of ${1/2}^+$ octet of baryons, unifying
compatibly all known properties of FF's, which renders just $\rho$-,
$\omega$-~and $\phi$-meson coupling constant ratios as free parameters.

For nucleons, these free parameters are evaluated numerically by a
comparison of the nucleon model with existing nucleon FF data and
then it is straightforward to find numerical values of $f_{\rho
NN}$, $f_{\omega NN}$ and $f_{\phi NN}$.
\begin{figure}[hbt]
\centerline{\epsfig{file=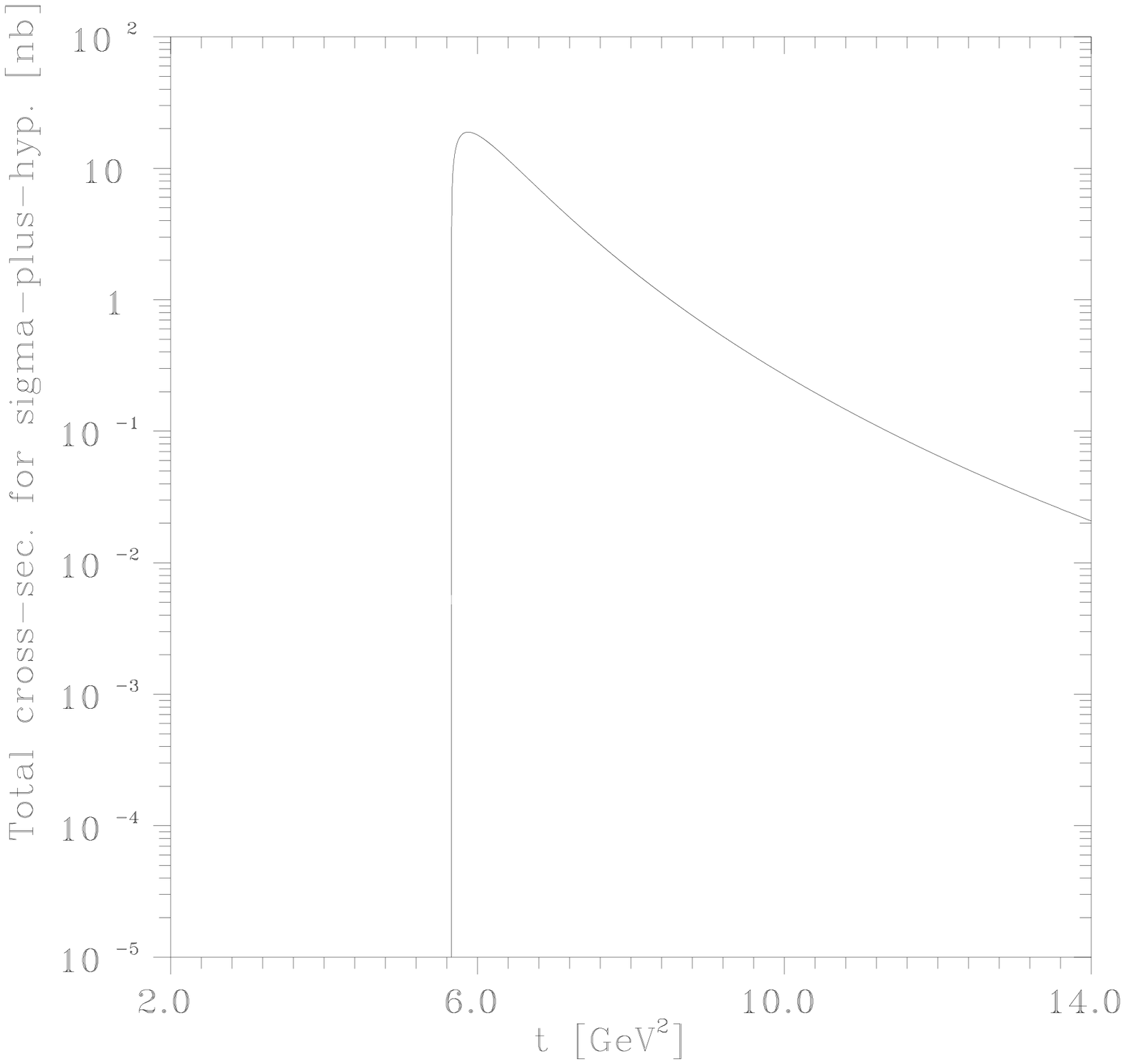,width=0.7\textwidth,
height=0.25\textheight}} \caption{}\label{fighyp2}
\end{figure}

On the other hand, the trace of SU(3) invariant Lagrangian for
vector-meson-baryon-antibaryon vertex
\begin{eqnarray}
{\rm Tr}(L_{VB\bar{B}}) &=&
\frac{i}{\sqrt{2}}f^F\left[\bar{B}^\alpha_\beta\gamma_\mu
B^\beta_\gamma-
\bar{B}^\beta_\gamma\gamma_\mu B^\alpha_\beta\right](V_\mu)^\gamma_\alpha + \nonumber\\
&+&\frac{i}{\sqrt{2}}f^D\left[\bar{B}^\beta_\gamma\gamma_\mu\beta^\alpha_\beta +
\bar{B}^\alpha_\beta\gamma_\mu B^\beta_\gamma\right](V_\mu)^\gamma_\alpha + \nonumber\\
&+&\frac{i}{\sqrt{2}}f^S\bar{B}^\alpha_\beta\gamma_\mu B^\beta_\alpha\omega^0_\mu
\end{eqnarray}
with $\omega-\phi$ mixing
\begin{eqnarray}
\phi^0 &=& \phi_8\cos\vartheta-\omega_1\sin\vartheta \nonumber \\
\omega^0 &=& \phi_8\sin\vartheta+\omega_1\cos\vartheta
\end{eqnarray}
$B$,$\bar{B}$ and $V$ baryon, antibaryon and vector-meson octuplet matrices,
$\omega^0_\mu$ omega-meson singlet, $f^F$, $f^D$ and $f^S$ SU(3) coupling constants
and $\vartheta$ mixing angle, provides the following expressions for
vector-meson-baryon coupling constants
\begin{eqnarray}
f^{(1,2)}_{\rho NN} &=& \frac{1}{2}(f^D_{1,2}+f^F_{1,2}) \nonumber \\
f^{(1,2)}_{\omega NN} &=&\frac{1}{\sqrt{2}}\cos\vartheta f^S_{1,2}-
      \frac{1}{2\sqrt{3}}\sin\vartheta(3f^F_{1,2}-f^D_{1,2}) \label{su3nuc}\\
f^{(1,2)}_{\phi NN} &=&\frac{1}{\sqrt{2}}\sin\vartheta f^S_{1,2} +
      \frac{1}{2\sqrt{3}}\cos\vartheta(3f^F_{1,2}-f^D_{1,2}) \nonumber
\end{eqnarray}

\begin{figure}[hbt]
\centerline{\epsfig{file=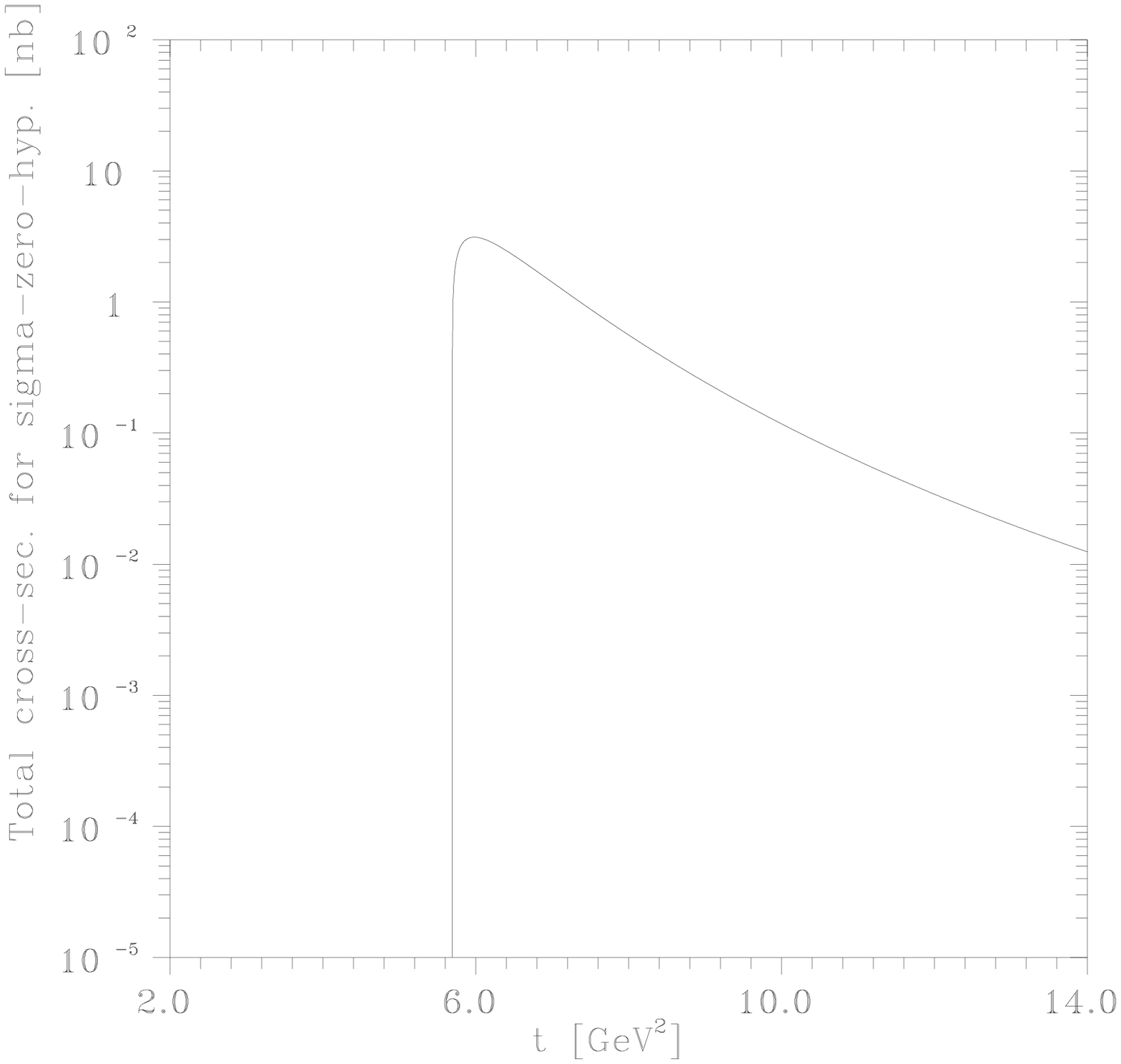,width=0.7\textwidth,
height=0.25\textheight}} \caption{}\label{fighyp3}
\end{figure}

\begin{eqnarray}
f^{(1,2)}_{\omega\Lambda\Lambda} &=& \frac{1}{\sqrt{2}}\cos\vartheta f^S_{1,2} +
          \frac{1}{\sqrt{3}}\sin\vartheta f^D_{1,2} \nonumber \\
 & & \label{su3lam}\\
f^{(1,2)}_{\phi\Lambda\Lambda} &=& \frac{1}{\sqrt{2}}\sin\vartheta f^S_{1,2} -
          \frac{1}{\sqrt{3}}\cos\vartheta f^D_{1,2} \nonumber
\end{eqnarray}

\begin{figure}[hbt]
\centerline{\epsfig{file=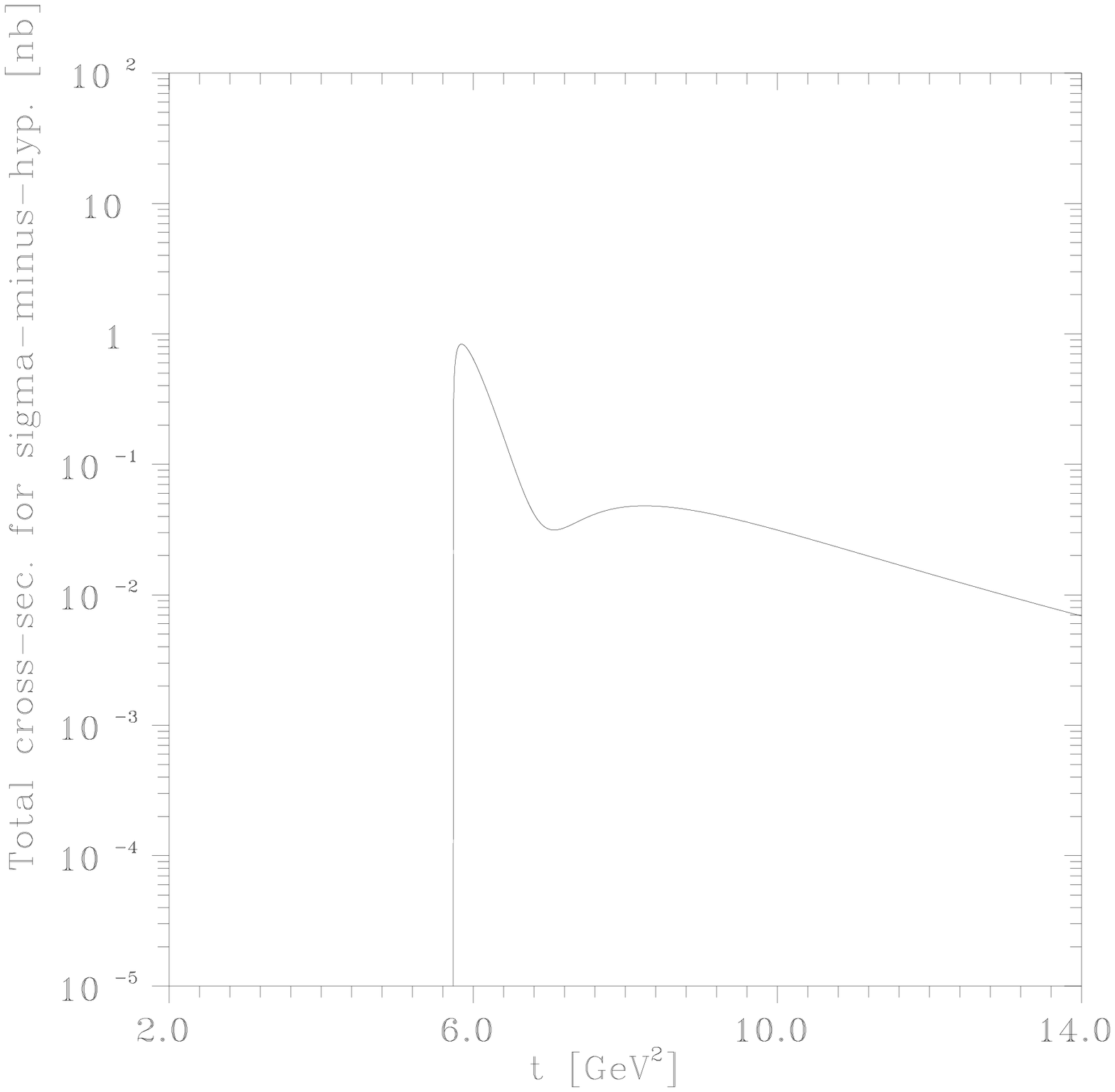,width=0.7\textwidth,
height=0.3\textheight}} \caption{}\label{fighyp4}
\end{figure}

\begin{eqnarray}
f^{(1,2)}_{\rho\Sigma\Sigma} &=& -f^F_{1,2} \nonumber \\
f^{(1,2)}_{\omega\Sigma\Sigma} &=& \frac{1}{\sqrt{2}}\cos\vartheta f^S_{1,2} -
         \frac{1}{\sqrt{3}}\sin\vartheta f^D_{1,2} \label{su3sig}\\
f^{(1,2)}_{\phi\Sigma\Sigma} &=& \frac{1}{\sqrt{2}}\sin\vartheta f^S_{1,2} +
         \frac{1}{\sqrt{3}}\cos\vartheta f^D_{1,2}\  \nonumber
\end{eqnarray}

\begin{figure}[hbt]
\centerline{\epsfig{file=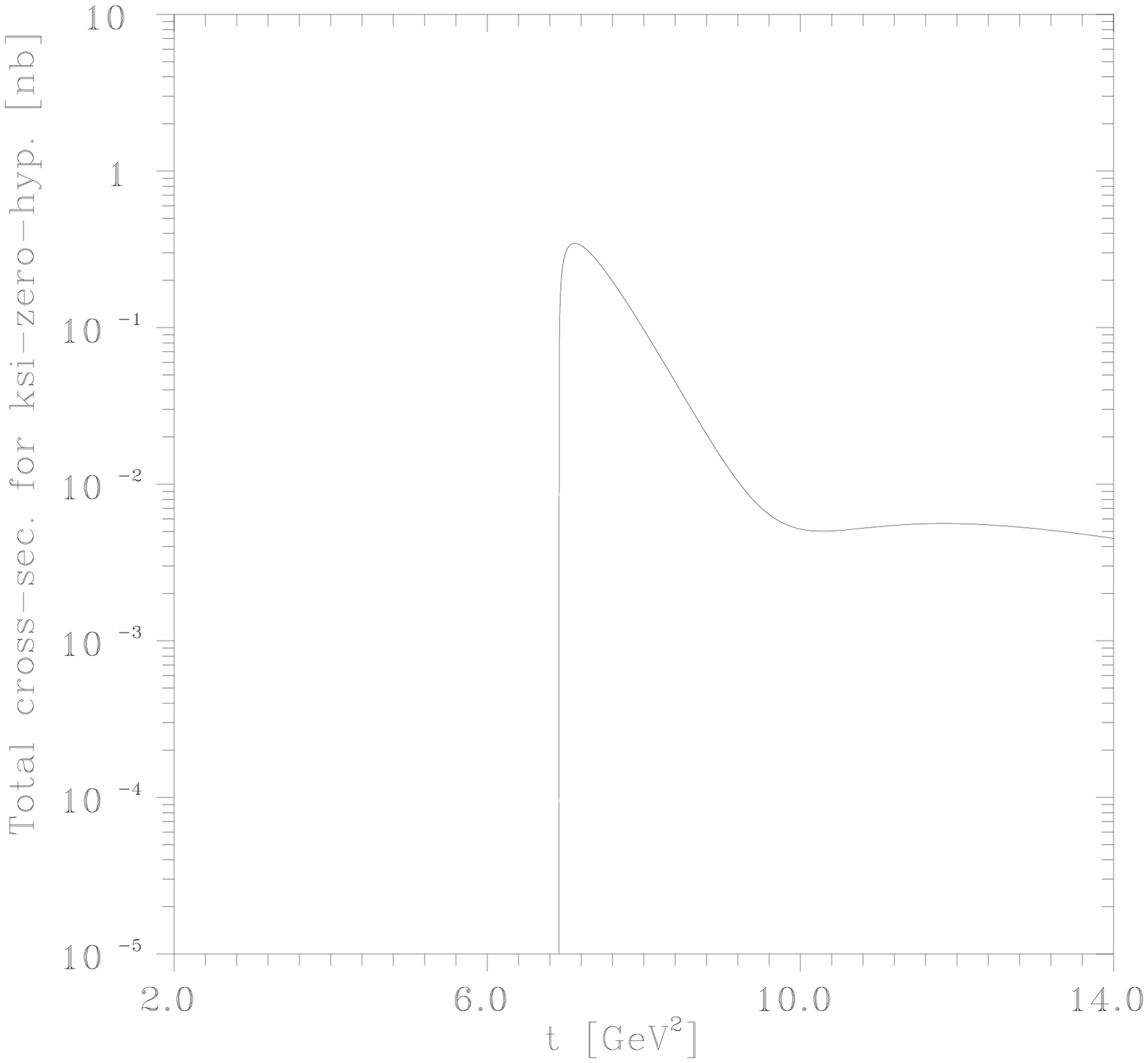,width=0.7\textwidth,
height=0.25\textheight}} \caption{}\label{fighyp5}
\end{figure}
\begin{figure}[hbt]
\centerline{\epsfig{file=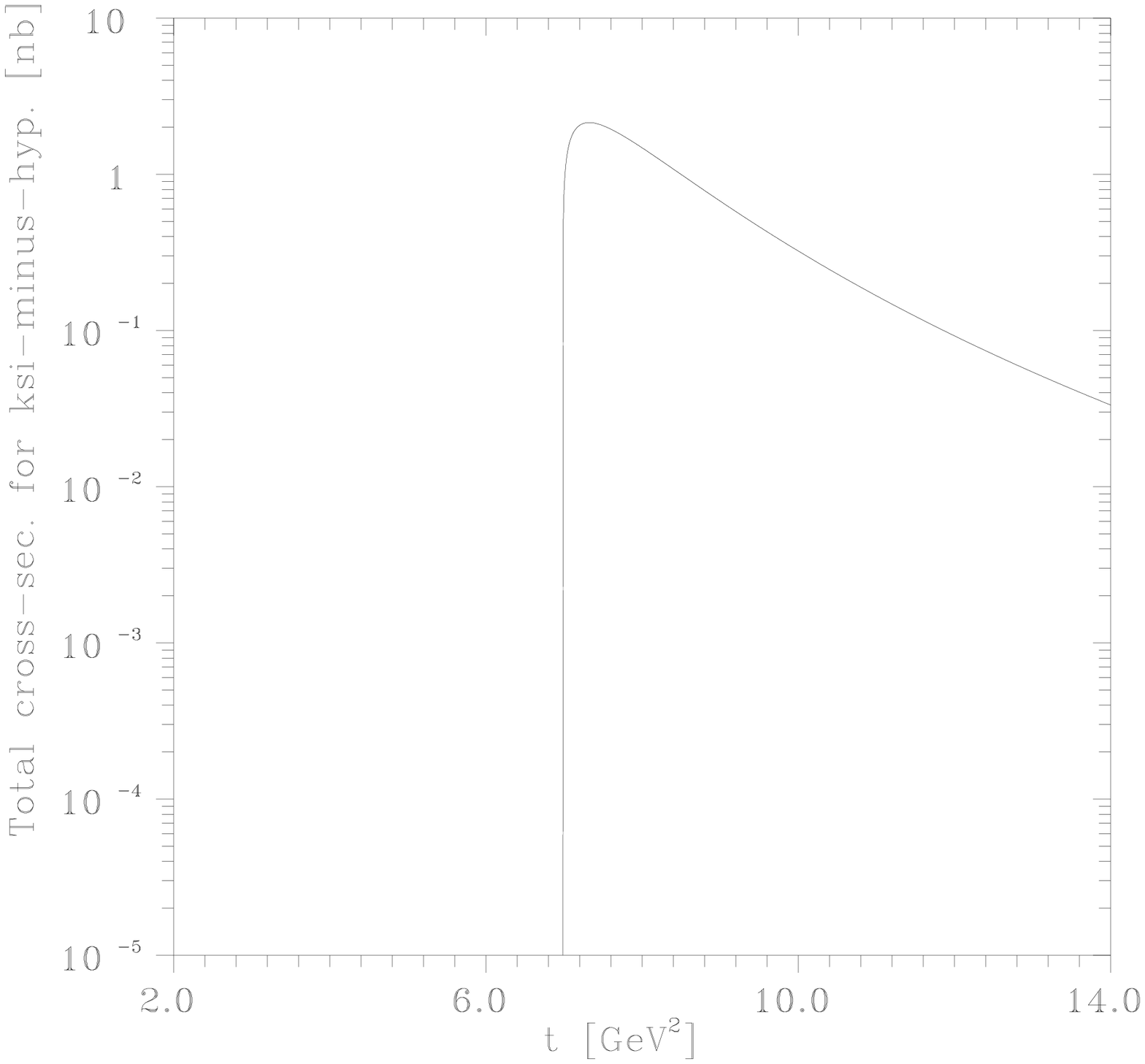,width=0.7\textwidth,
height=0.25\textheight}} \caption{}\label{fighyp6}
\end{figure}


\begin{eqnarray}
f^{(1,2)}_{\rho\Xi\Xi} &=& \frac{1}{2}(f^D_{1,2}-f^F_{1,2}) \nonumber \\
f^{(1,2)}_{\omega\Xi\Xi} &=&\frac{1}{\sqrt{2}}\cos\vartheta f^S_{1,2}+
      \frac{1}{2\sqrt{3}}\sin\vartheta(3f^F_{1,2}+f^D_{1,2}) \label{su3xi}\\
f^{(1,2)}_{\phi\Xi\Xi} &=&\frac{1}{\sqrt{2}}\sin\vartheta f^S_{1,2} -
      \frac{1}{2\sqrt{3}}\cos\vartheta(3f^F_{1,2}+f^D_{1,2}). \nonumber
\end{eqnarray}
Then the solutions of the system of algebraic eqs.~(\ref{su3nuc})
according to $f^D_{1,2}$, $f^F_{1,2}$, $f^S_{1,2}$ with numerical
values of $f_{\rho NN}^{1,2}$, $f_{\omega NN}^{1,2}$ and $f_{\phi
NN}^{1,2}$, enable by means of the expressions
(\ref{su3lam})-(\ref{su3xi})to predict all free
vector-meson-hyperon coupling constant ratios in the \emg~FF's of
hyperons $[\Lambda^0]$, $[\Sigma^+,\Sigma^0,\Sigma^-]$ and
$[\Xi^0,\Xi^-]$ and, as a result, also behaviours of the total
cross-sections (\ref{stothyp}) (see
Figs.~\ref{fighyp1}-\ref{fighyp6}).

\subsection{Estimation of hadronic contributions to the anomalous magnetic
moment of the muon with higher precision}

Charged leptons $(l=e^-,\mu^-,\tau^-)$ are described by the Dirac equation and their
magnetic moments are related to the spin by the relation
\begin{equation}
\mu_l=g_l\;\frac{e}{2m_l\cdot c}\;\frac{\hbar}{2}
\end{equation}
where the value of the gyromagnetic ratio $g_l$ is predicted (in the absence
of the Pauli term) to be exactly 2.

Practically, however, the interactions existing in nature modify
$g_l$ to be different from 2 because of the emission and absorption
of virtual photons (\emg~effects), intermediate vector and Higgs
bosons (weak interaction effects) and the vacuum polarization into
virtual hadronic states (strong interaction effects).

In order to describe theoretically this modification of $g_l$, the magnetic
anomaly was introduced by the relation
\begin{equation}
a_l\equiv\frac{g_l-2}{2}=a^{(1)}_l\cdot\frac{\alpha}{\pi}+
(a^{(2)\;QED}_l+a^{(2)\;had}_l)\cdot\left(\frac{\alpha}{\pi}\right)^2+
a^{(2)\;weak}_l+O(\frac{\alpha}{\pi})^3
\end{equation}
where $\alpha=1/137.0359895(61)$ is the fine-structure constant.

From all three leptons the most interesting for theoretical investigations
is the muon magnetic anomaly $a_\mu$:
\begin{itemize}
\item[$\bullet$]
it is one of the best measured quantities in physics
\begin{equation}
a^{\rm exp}_\mu = (116\;592\;290\;\pm\;850)\times 10^{-11}\;\;\;\;\cite{ammexp}
\label{ammcern}
\end{equation}
\item[$\bullet\bullet$]
its accurate theoretical evaluation provides an extremely clean test of "Electroweak
theory" and may give hints on possible deviations from STANDARD MODEL (SM).
\item[$\bullet\bullet\bullet$]
the NEW E-821 experiment~\cite{ammexpnew} in Brookhaven National Laboratory (BNL)
is under way, which is expected to be performed with an accuracy
\begin{equation}
\Delta a^{\rm exp}_\mu =\;\pm\;43\;\times 10^{-11}.
\label{newerr}
\end{equation}
i.e. it is aimed at obtaining a factor of 20 in a precision above that of the
previous CERN measurements~(\ref{ammcern}).
\end{itemize}

Comparing theoretical evaluations of QED, WEAK and STRONG effects
\begin{itemize}
\item[]
QED:\\
\begin{eqnarray}
a^{\rm QED}_\mu &=& \;(116584694.7\;\pm\;4.6)\times 10^{-11}
\hbox{\rm\mbox{~} \hfill (KNO)~\cite{kin1}}\nonumber \\
 & & \label{ammthqed} \\
a^{\rm QED}_\mu &=& \;(116584698.4\;\pm\;1.7)\times 10^{-11}
\hbox{\rm\mbox{~} \hfill (K)~\cite{kin2}}\nonumber
\end{eqnarray}
\item[]
WEAK:\\
\begin{eqnarray}
a^{\rm W}_\mu &=& \;(155.1\;\pm\;2.7)\times 10^{-11}
\hbox{\rm\mbox{~} \hfill (PPdeR)~\cite{ppr}}\nonumber \\
 & & \label{ammthw} \\
a^{\rm W}_\mu &=& \;(152\;\pm\;3\;\pm\;0.45{\rm R}_{\rm b})\times
10^{-11} \hbox{\rm\mbox{~} \hfill (CzKM)~\cite{czkm}}\nonumber
\end{eqnarray}
\item[]
STRONG:\\
\begin{eqnarray}
a^{\rm had}_\mu &=& \;(7068\;\pm\;172)\times 10^{-11}
\hbox{\rm\mbox{~} \hfill (KNO)~\cite{kin3}}\nonumber \\
a^{\rm had}_\mu &=& \;(7100\;\pm\;116)\times 10^{-11}
\hbox{\rm\mbox{~} \hfill (CLY)~\cite{cly}}\label{ammths}\\
a^{\rm had}_\mu &=& \;(7024\;\pm\;152)\times 10^{-11}
\hbox{\rm\mbox{~} \hfill (EJ)~\cite{ej}}\nonumber
\end{eqnarray}
\end{itemize}
it is straightforward to see, that the largest uncertainty comes from STRONG
interaction contributions~(\ref{ammths}). Their errors are comparable
with central values of WEAK interaction effects in~(\ref{ammthw}).

So, if one would like to test the SM predictions for $a^W_\mu$, possibly to obtain
hints for physics beyond SM, one has still to lower the error of $a^{\rm had}_\mu$.

The latter quantity can be represented by the integral
\begin{equation}
a^{(2)\;\rm had}_\mu = \frac{1}{4\pi^3}\int^\infty_{4m^2_\pi}
\sigma^h(s)K^{(2)}_\mu(s){\rm d}s
\label{ammhad}
\end{equation}
where
\begin{eqnarray}
\sigma^h(s)\equiv\sigma_{\rm tot}(e^+e^-\to {\rm had}) &=&
\sigma_{\rm tot}(e^+e^-\to\pi^+\pi^-) + \sigma_{\rm tot}(e^+e^-\to K^+K^-) + \nonumber \\
& & \label{ammdec}\\
&+&\sigma_{\rm tot}(e^+e^-\to K^0\bar{K^0}) + \sigma_{\rm tot}(e^+e^-\to\pi^+\pi^-\pi^0) + \cdots \nonumber
\end{eqnarray}
and
\begin{equation}
K^{(2)}_\mu(s) = \int^1_0\frac{x^2(1-x)}{x^2+(1-x)s/m^2_\mu}{\rm d}x .
\end{equation}

From the practical point of view it was advantageous to separate the low energy
$(4m^2_\pi < s < s_0)$ from the high energy region $(s_0 < s < \infty)$ in the
integral~(\ref{ammhad}) as follows
\begin{eqnarray}
a^{(2)\;\rm had}_\mu = \frac{1}{4\pi^3}&&
\left\{\int^{s_0}_{4m^2_\pi}\left[\sum_F\sigma_{\rm tot}(e^+e^-\to F)\right]
K^{(2)}_\mu(s){\rm d}s + \right.\\
&&\Biggl. +\int^\infty_{s_0}R(e^+e^-\to {\rm had})\sigma_{\rm tot}(e^+e^-\to\mu^+\mu^-)K^{(2)}_\mu(s){\rm d}s \Biggr\}.\nonumber
\end{eqnarray}

The $s_0$ was fixed at 2~GeV$^2$, where an overlapping region of
the data on total cross-sections of exclusive processes $e^+e^-\to
F$ and the data on $R= \sigma_{\rm tot}(e^+e^-\to {\rm
had})/\sigma_{\rm tot}(e^+e^-\to\mu^+\mu^-)$ already exists.

Then the first error lowering in $a^{\rm had}_\mu$ was achieved in~\cite{amm1}
\begin{equation}
a^{\rm had}_\mu =\;(7052\;\pm\;76)\times 10^{-11}
\end{equation}
where instead of an integration over the experimental data on the first three
total cross-sections in~(\ref{ammdec}) refined models of the corresponding FF's
were used and for a determination of their free parameters not only the data
of the region of a definition of $\sigma_{\rm tot}(e^+e^-\to P\bar{P})\;\;(P=\pi, K)$,
but all existing space-like and time-like data, were used.

Even better error lowering in  $a^{\rm had}_\mu$ was reached in~\cite{amm2}
\begin{equation}
a^{\rm had}_\mu =\;(6986\;\pm\;45)\times 10^{-11}
\end{equation}
if for the corresponding FF's the sophisticated unitary and analytic models,
discussed in Sect.~\ref{secmainidea}, were applied. Really, in this case the error
is diminished down to one fourth of the WEAK interaction contribution~(\ref{ammthw})
and becomes comparable with the accuracy~(\ref{newerr}) expected in the new $g-2$
muon running experiment at BNL~\cite{ammexpnew}.

\section{Conclusions}

We have formulated main principles of the unitary and analytic
models of the electromagnetic structure of hadrons and nuclei. Then
a general scheme of their utilization to the \emg, weak and strong
interaction processes was traced out. Finally, their practical
application for a description of the \emg~and weak FF's, by means
of which a lot of interesting results were obtained, convinces us
that they are still powerful tools in the particle physics
phenomenology.

\section{Acknowledgements}

The work was partly supported by Slovak Grant Agency for Sciences,
Grant No. 2/5085/2000 (S.D. and P.S.) and Grant No. 1/7068/2000
(A.Z.D.).

\end{document}